\documentclass[%
 aip,
 amsmath,amssymb,
 preprint,%
]{revtex4-1}

\usepackage[utf8]{inputenc}
\usepackage[T1]{fontenc}
\usepackage{mathptmx}
\usepackage{mathtools}
\usepackage{hyperref}
\usepackage[inline]{enumitem}

\DeclareUnicodeCharacter{2009}{\,} 
\DeclareUnicodeCharacter{03C0}{$\pi$} 

\usepackage{graphicx}
\usepackage{physics}
\usepackage{makecell}
\usepackage{braket}
\usepackage{amsmath}
\usepackage[T1]{fontenc} 
\newcommand{\RomanNumeralCaps}[1]
    {\MakeUppercase{\romannumeral #1}}

\begin{document}

\preprint{AIP/123-QED}

\title{Numerical Assessment for Accuracy and GPU Acceleration of TD-DMRG Time Evolution Schemes}

\author{Weitang Li}
 \affiliation{MOE Key Laboratory of Organic OptoElectronics and Molecular
 Engineering, Department of Chemistry, Tsinghua University, Beijing 100084,
 People's Republic of China }
 
\author{Jiajun Ren}
\email{renjj@mail.tsinghua.edu.cn}
 \affiliation{MOE Key Laboratory of Organic OptoElectronics and Molecular
 Engineering, Department of Chemistry, Tsinghua University, Beijing 100084, People's Republic of China }

\author{Zhigang Shuai}%
 \affiliation{MOE Key Laboratory of Organic OptoElectronics and Molecular
 Engineering, Department of Chemistry, Tsinghua University, Beijing 100084,
 People's Republic of China }
 
\date{\today}

%

\begin{abstract}
Time dependent density matrix renormalization group (TD-DMRG) has become one of the cutting edge methods
of quantum dynamics for complex systems. 
In this paper, we comparatively study the accuracy of three
time evolution schemes in TD-DMRG, the global propagation and compression method with Runge-Kutta algorithm (P\&C-RK), 
the time dependent variational principle based methods with matrix unfolding algorithm (TDVP-MU) and with projector-splitting algorithm (TDVP-PS),
by performing benchmarks on the exciton dynamics of Fenna-Matthews-Olson (FMO) complex.
We show that TDVP-MU and TDVP-PS yield the same result when the time step size is converged and they are more accurate than P\&C-RK4, while TDVP-PS  tolerates a larger time step size than TDVP-MU.
We further adopt the graphical processing units (GPU) to accelerate the heavy tensor contractions in TD-DMRG
and it is able to speed up the TDVP-MU and TDVP-PS schemes by up to 73 times.
\end{abstract}

\maketitle

\section{Introduction}
Time dependent density matrix renormalization group (TD-DMRG) has emerged as a powerful tool to deal with 
many-body chemical and physical problems~\cite{ma2018time,paeckel2019time}.
Although DMRG is initially designed to solve the ground state of one dimensional strongly correlated systems~\cite{white1992density,white1993density}, 
its applications are successfully extended to dynamical properties both in the time and frequency domains, such as
linear and nonlinear optical response of polyenes~\cite{shuai1998linear},
polaron formation and diffusion~\cite{zhang1999dynamical,yao2017polaronic,barford2018torsionally},
interconversion dynamics of pyrazine~\cite{greene2017tensor,baiardi2019large,xie2019time},
ab initio electron dynamics in hydrogen chain~\cite{ronca2017time},
exciton dissociation\cite{yao2018full},
spectra of molecular aggregates~\cite{ren2018time} and
many other topics~\cite{chin2013role,schroder2019tensor,borrelli2017simulation,kloss2019spin,kloss2019multiset,frahm2019ultrafast}.
One of the key components in TD-DMRG is the time evolution scheme, which is essential to the numerical accuracy and efficiency. The available schemes could be roughly classified into three groups. 
The first group is based on globally approximating the formal propagator $e^{-iHt}$ or $e^{-iHt}\ket{\Psi}$, including time-evolving block decimation (TEBD)~\cite{vidal2004efficient,white2004real,daley2004time}, W\textsuperscript{\RomanNumeralCaps{1},\RomanNumeralCaps{2}} method~\cite{zaletel2015time}, Runge-Kutta~\cite{garcia2006time,ren2018time}, Chebyshev expansion~\cite{halimeh2015chebyshev}, Krylov subspace ~\cite{garcia2006time,wall2012out} methods and split operator method on the grid basis~\cite{greene2017tensor}. The same feature shared in these schemes is that in each time step the wavefunction is firstly propagated as a whole globally and then compressed. 
The second group is more inspired by the original DMRG, which is formulated in the local renormalized space and the basis is adapted by the averaged reduced density matrix. The representatives are time step targeting method (TST) ~\cite{feiguin2005time} and some related variants~\cite{dutta2010double,ronca2017time}. 
The third group is based on the time dependent variational principle (TDVP)~\cite{dirac1930note}. Depending on the different ways to derive the equations of motion (EOMs), this group includes original method with fixed gauge freedom ~\cite{haegeman2011time} and the more recent projector-splitting method (PS) from a tangent space view ~\cite{haegeman2016unifying}.
Among the above evolution schemes, all schemes are suited to long-range interactions except TEBD. In addition, the global evolution scheme is the most straightforward one when the modern framework of matrix product state / matrix product operator (MPS/MPO) is investigated,
while the PS scheme seems to have become the most popular choice 
as it has been widely employed in the recent articles~\cite{borrelli2017simulation,schroder2019tensor, baiardi2019large, kloss2019spin,kloss2019multiset,kurashige2018matrix,xie2019time}
and implemented in a number of TD-DMRG packages~\cite{hauschild2018efficient,mendl2018pytenet}.
Although many evolution schemes have been applied extensively to the simulation of quantum dynamics in both chemistry and physics, 
a pragmatic analysis of their relative accuracy and efficiency has not been reported yet.

High performance computing for DMRG algorithms has attracted much interest in recent years, including parallelization strategies with message passing interface (MPI)~\cite{chan2004algorithm}, open multi-processing (OpenMP) and hybrid MPI/OpenMP\cite{kurashige2009high} on multiple central processing units (CPUs). Meanwhile, the application of graphical processing units (GPUs) in computational chemistry has also drawn much attention in the past decade, 
due to the tardy improvement of CPUs,  
and the more and more vibrant GPU software ecosystem, including electronic structure calculation~\cite{song2017analytical,kaliman2017new,shee2018phaseless,liu2019exploiting}, classical/ab initio molecular dynamics~\cite{penfold2017accelerating,lee2018gpu}, and open system quantum dynamics~\cite{kreisbeck2011high,tsuchimoto2015spins}. 
To the best of our knowledge, the only attempt to employ GPUs 
in DMRG was made by Nemes \textit{et al}. in 2014~\cite{nemes2014density}. 
They came up with a smart implementation exploiting both CPU and GPU which speeds up the Davidson diagonalization part in
DMRG algorithm by 2 to 5 times.
Despite their efforts, how GPUs can accelerate TD-DMRG algorithms and which evolution scheme is more suitable remain unclear.

To benchmark the performance of different TD-DMRG time evolution schemes, a proper model should be chosen. In this work, we focus on the vibronic coupling system represented by the Fenna-Matthews-Olson (FMO) complex from green sulfur bacteria.
It is an extremely popular system for both experimental and theoretical studies of 
energy transfer in photosynthesis~\cite{engel2007evidence,cheng2009dynamics}. 
The FMO model has become a “guinea pig” for comparing different computational methods~\cite{borrelli2017simulation}
and its exciton dynamics has been studied by many numerically exact methods, including
quasi-adiabatic propagator path integral (QUAPI)~\cite{nalbach2011exciton},
hierarchical equations of motion (HEOM)~\cite{ishizaki2009theoretical,shi2018efficient},
multi-layer multi-configuration time-dependent Hartree (ML-MCTDH)~\cite{schulze2015explicit,schulze2016multi},
TD-DMRG~\cite{chin2013role,borrelli2017simulation}, etc.

In this paper, we select three time evolution schemes in TD-DMRG to simulate the exciton dynamics of 7-site FMO model, including global propagation and compression scheme with classical 4th order Runge-Kutta algorithm (P\&C-RK4), TDVP with advanced matrix unfolding regularization scheme (TDVP-MU) 
and TDVP with projector-splitting algorithm (TDVP-PS). 
We firstly study the relative accuracy of the three schemes on the FMO model and then set out to explore the CPU-GPU heterogeneous computing to accelerate the TD-DMRG algorithms.

\section{Methodological Approaches}
\label{sec:methods}

\newcommand{\dt}{\tau}

\subsection{MPS Representation and TD-DMRG Algorithms}
\begin{figure}
  \includegraphics[width=.48\textwidth]{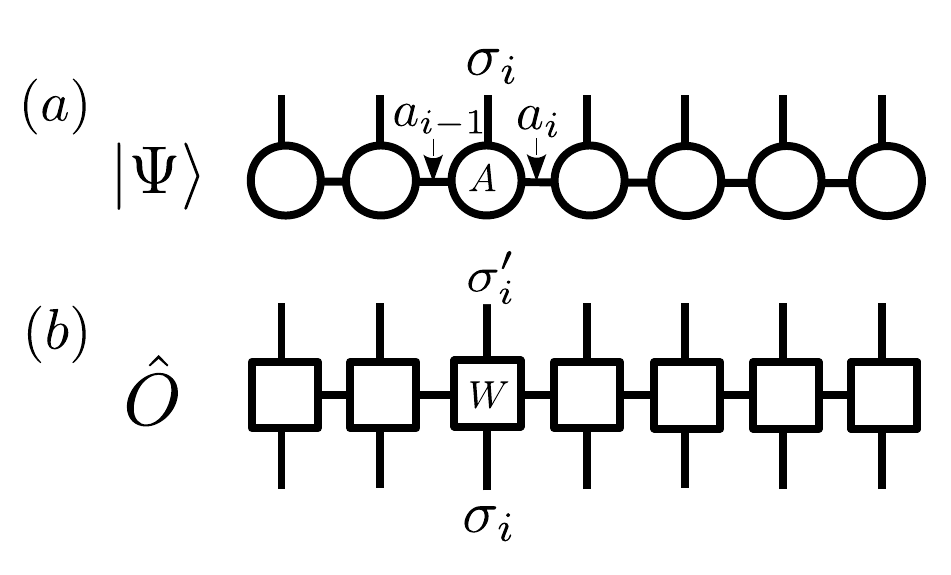}
  \caption{The graphical representation of (a) an MPS in Eq.~\ref{eq:mps} (b) an MPO in Eq.~\ref{eq:mpo} with $N=7$.}
  \label{fig:psioperator}
\end{figure}

In the language of matrix product states (MPS)~\cite{schollwock2011density}, a quantum state $\ket{\Psi}$
under certain basis $\ket{ \sigma_1\sigma_2\cdots\sigma_N } $, where $\ket{\sigma_i}$ is the basis for each degree of freedom (DOF) and is assumed to be orthonormal, can be represented as the product of a matrix chain,
known as an MPS:

\begin{equation}
\label{eq:mps}
    \ket{\Psi}  = \sum_{\{a\},\{\sigma\}}
     A^{\sigma_1}_{a_1} A^{\sigma_2}_{a_1a_2} \cdots
           A^{\sigma_N}_{a_{N-1}}  \ket{ \sigma_1\sigma_2\cdots\sigma_N }    
\end{equation}
$A^{\sigma_i}_{a_{i-1}a_{i}}$ are matrices in the chain connected by indices $a_i$. $\{ \cdot \}$ in the summation represents the contraction of the respective connected indices, 
and $N$ is the total number of DOFs in the system.
The dimension of $a_i$ is called (virtual) bond dimension denoted as $M_{\textrm{S}}$ or $|a_i|$,
while the dimension of $\sigma_i$ is called physical bond dimension denoted as $d$. The graphical representation of an MPS is shown in Fig.~\ref{fig:psioperator}(a).
The many-body renormalized basis is defined as:
\begin{align}
\label{eq:renormalized-basis}
    \ket{a_{i}[1:i]} & = \sum_{\{a\},\{\sigma\}} 
        A^{\sigma_1}_{a_1} A^{\sigma_2}_{a_1a_2} \cdots A^{\sigma_{i}}_{a_{i-1}a_{i}} \ket{ \sigma_{1} \cdots\sigma_{i} }   \\
    \ket{a_{j}[j+1:N]} & = \sum_{\{a\},\{\sigma\}}  
     A^{\sigma_{j+1}}_{a_j a_{j+1}} \cdots
           A^{\sigma_N}_{a_{N-1}}  \ket{ \sigma_{j+1} \cdots\sigma_N }
\end{align}
where $[1:i]$ and $[j+1:N]$ represent that the renormalized bases are defined 
from the left and right side respectively. 
Hence, $A^{\sigma_n}_{a_{n-1}a_n}$ could be regarded as the coefficient matrix on the space spanned by $\ket{a_{n-1}[1:n-1]} \otimes \ket{\sigma_n}  \otimes \ket{a_{n}[n+1:N]}$. However, in general, the renormalized basis is not necessary to be orthonormal, and the overlap matrix between them is defined as:
\begin{gather}
    S[1:i]_{a_{i}a'_{i}} =  \langle a_{i}[1:i] |a'_{i}[1:i]\rangle  \\
    S[j+1:N]_{a_{j} a'_{j}} =  \langle a_{j}[j+1:N] |a'_{j}[j+1:N]\rangle 
\end{gather}

The matrix product representation for a wavefunction is not unique in that inserting an identity matrix $I=GG^{-1}$ into the neighboring matrices $\cdots A^{\sigma_i}A^{\sigma_{i+1}}\cdots$ will obtain the same wavefunction but with different local matrices $\cdots A'^{\sigma_i}A'^{\sigma_{i+1}}\cdots=\cdots(A^{\sigma_i}G)(G^{-1}A^{\sigma_{i+1}})\cdots=\cdots A^{\sigma_i}A^{\sigma_{i+1}}\cdots$. Therefore, gauge conditions could be applied to eliminate the parameterization redundancy of an MPS. 
Among them, ``mixed/left/right-canonical'' gauge condition is usually adopted for convenience. 
A ``mixed-canonical'' MPS with gauge center at site $n$ is written as:
\begin{equation}
\begin{aligned}
\label{eq:mix-canon}
\ket{\Psi} = \sum_{\{l\},\{r\}\{\sigma\}} &
     L^{\sigma_1}_{l_1} L^{\sigma_2}_{l_1l_2} \cdots L^{\sigma_{n-1}}_{l_{n-2}l_{n-1}}C^{\sigma_n}_{l_{n-1}r_n}
     R^{\sigma_{n+1}}_{r_n r_{n+1}} \cdots
           R^{\sigma_N}_{r_{N-1}} \ket{ \sigma_1\sigma_2\cdots\sigma_N }
\end{aligned}
\end{equation}
where $L^{\sigma_i}_{l_{i-1}l_{i}}$ and $R^{\sigma_j}_{r_{j-1}r_{j}}$ satisfy:
\begin{align}
    \sum_{\sigma_i, l_{i-1}}L^{\sigma_i*}_{l_{i-1}l'_{i}} L^{\sigma_i}_{l_{i-1}l_{i}} & = \delta_{l'_i l_i} \label{eq:lcanon}\\
    \sum_{\sigma_j, r_{j}}R^{\sigma_j*}_{r'_{j-1}r_{j}} R^{\sigma_j}_{r_{j-1}r_{j}} & = \delta_{r'_{j-1} r_{j-1}} \label{eq:rcanon}
\end{align}
here $L^{\sigma_i}(R^{\sigma_j})^*$ represents the conjugate of $L^{\sigma_i}(R^{\sigma_j})$. 

With Eq.~\ref{eq:lcanon} and ~\ref{eq:rcanon}, the renormalized basis fulfill the orthonormal relation $  S[1:i]_{l_{i} l'_{i}}  = \delta_{l_{i} l'_{i}}$ ($i=1,2,\cdots,n-1$) and 
$ S[j+1:N]_{r_{j}  r'_{j}}  = \delta_{r_{j}  r'_{j}}$ ($j=n,n+1,\cdots,N-1$). 
When the gauge center $n$ is at the right(left)-most site, the MPS is called left(right)-canonical state. 
With the `mixed/left/right-canonical'' condition, the only redundancy left is that G is unitary. 
If we further decompose $C^{\sigma_n}$ by QR decomposition
\begin{gather}
    C^{\sigma_n}_{l_{n-1} r_n} \overset{\textrm{reshape}}{\to}  C_{\sigma_n l_{n-1}, r_n}  \overset{\textrm{QR}}{=} \sum_{l_n} L_{\sigma_n l_{n-1}, l_n} D_{l_n, r_n}  \overset{\textrm{reshape}}{\to} \sum_{l_n} L^{\sigma_n}_{l_{n-1} l_n} D_{l_n r_n} \label{eq:QR}
\end{gather}
$L^{\sigma_n}$ fulfills the relation in Eq.~\ref{eq:lcanon} and $D_{l_n r_n}$ is also a coefficient matrix defined between site $n$ and $n+1$ on space $\ket{l_n[1:n]} \otimes \ket{r_n[n+1:N]}$. Afterwards, $D_{l_n r_n}$ is combined with $R^{\sigma_{n+1}}_{r_n r_{n+1}}$ to obtain $C^{\sigma_{n+1}}_{l_{n} r_{n+1}} = \sum_{r_n} D_{l_n r_n} R^{\sigma_{n+1}}_{r_n r_{n+1}}$ and apparently the gauge center is moved one site to the right. The reverse process to move the gauge center to the left could be carried out in a similar way as Eq.~\ref{eq:QR} while QR is replaced with RQ decomposition.
More generally, a canonical MPS with gauge center at $n$ as Eq.~\ref{eq:mix-canon} could be prepared by preforming QR decomposition from site 1 to $n-1$ sequentially and RQ decomposition from site $N$ to $n+1$ sequentially on any MPS, which is called canonicalisation. 
In the following paper, QR and RQ are not distinguished for simplicity.

If the QR decomposition in Eq.~\ref{eq:QR} is replaced with singular value decomposition (SVD),
\begin{gather}
    C^{\sigma_n}_{l_{n-1} r_n} \overset{\textrm{reshape}}{\to}  C_{\sigma_n l_{n-1}, r_n}  \overset{\textrm{SVD}}{=} \sum_{s=1}^{k} U_{\sigma_n l_{n-1}, s} \Lambda_{ss} V^\dagger_{s, r_n}  \overset{\textrm{reshape}}{\to} \sum_{s=1}^{k} U^{\sigma_n}_{l_{n-1} s} \Lambda_{ss} V^\dagger_{s r_n} \approx \sum_{l_n=1}^{M_{\textrm{S}}<k} L^{\sigma_n}_{l_{n-1} l_n} \Lambda_{l_n l_n} V^\dagger_{l_n r_n}  \label{eq:SVD}
\end{gather}
$\Lambda$ is a real diagonal matrix with elements ordered from large to small $\Lambda_{11} \geq \Lambda_{22} > \cdots \geq \Lambda_{kk} \geq 0$ ( $k=\min{[|\sigma_n|\cdot|l_{n-1}|, |r_{n}|]}$) and $\sum_s \Lambda^2_{ss}=1$ if $\ket{\Psi}$ is normalized. $U^{\sigma_n}$ fulfills the relation in Eq.~\ref{eq:lcanon}. If we retain all $\Lambda_{ss}$ ($M_\textrm{S}=k$) and move the gauge center to the right with $C^{\sigma_{n+1}}_{l_{n} r_{n+1}} = \sum_{l_n} \Lambda_{l_n l_n} V^\dagger_{l_n r_n} R^{\sigma_{n+1}}_{r_n r_{n+1}}$, the wavefunction is not changed and SVD plays the same role as QR in canonicalisation but with a little bit higher cost. 
However, if we only retain the largest $M_\textrm{S}$ terms ($M_\textrm{S}<k$) to obtain  $\tilde{C}^{\sigma_{n+1}}_{l_{n} r_{n+1}}$,  $\tilde{C}^{\sigma_{n+1}}_{l_{n} r_{n+1}}$ could be a good approximation to $C^{\sigma_{n+1}}_{l_{n} r_{n+1}}$ with a smaller bond dimension.
With this algorithm, a left(right)-canonical MPS $\ket{\Psi}$ could be ``compressed'' to $\ket{\Tilde{\Psi}}$ by successive approximate SVD decomposition from site $N$ to 1 (site 1 to $N$).
At each local step of the compression, two truncation criteria are commonly used
\begin{enumerate*}[label=(\roman*)]
\item a fixed pre-defined $M_\textrm{S}$;
\item adaptive $M_\textrm{S}$ with all $\Lambda_{ss}$ larger than pre-defined $\zeta$ retained.
\end{enumerate*}

Apart from canonicalisation and compression, several other operations are usually met within the MPS algorithms,
including $\hat{O}\ket{\Psi}$  and $\ket{\Psi} + \ket{\Phi}$. 
Similar to MPS, a common quantum operator could be expressed as a matrix product operator (MPO)~\cite{schollwock2011density,chan2016matrix}:
\begin{equation}
\label{eq:mpo}
    \hat{O} = \sum_{\{w\},\{\sigma\},\{\sigma'\}}
     W^{\sigma'_1, \sigma_1}_{w_1} W^{\sigma'_2, \sigma_2}_{w_1w_2} \cdots
                    W^{\sigma'_N, \sigma_N}_{w_{N-1}} 
                    \ket{\sigma'_1\sigma'_2\cdots\sigma'_N}
                    \bra{\sigma_N\sigma_{N-1} \cdots \sigma_1}
\end{equation}
The graphical representation of an MPO is shown in Fig.~\ref{fig:psioperator}(b).
With the local matrix representation of wavefunction and operator in Eq.~\ref{eq:mps} and Eq.~\ref{eq:mpo},
$\hat{O}\ket{\Psi}$ (MPO$\times$MPS) could be calculated by contracting $W^{\sigma'_i,\sigma_i} A^{\sigma_i}$ locally: 
\begin{gather}
    \hat{O} \ket{\Psi} = \sum_{\{w, a\},\{\sigma\}}
           A'^{\sigma'_1}_{\{w, a\}_1} A'^{\sigma'_2}_{\{w, a\}_1\{w, a\}_2} \cdots
           A'^{\sigma'_N}_{\{w, a\}_{N-1}}  \ket{ \sigma'_1\sigma'_2\cdots\sigma'_N }  \\
\shortintertext{where}
A'^{\sigma'_i}_{\{w, a\}_{i-1}\{w, a\}_{i}} = 
\sum_{\sigma_i} W^{\sigma'_i, \sigma_i}_{w_{i-1}w_{i}} A^{\sigma_i}_{a_{i-1}a_{i}} \label{eq:mpo-apply}
\end{gather}
Suppose the bond dimensions of the original MPS and the MPO are $M_\textrm{S}$ and $M_\textrm{O}$ respectively,
then the new state $\ket{\hat{O}  \Psi}$ has bond dimension $M_\textrm{O}M_\textrm{S}$. 
$\ket{\Psi} + \ket{\Phi}$ (MPS + MPS) is constructed by merging the local matrices $[A^{\sigma_i}, B^{\sigma_i}]$ block-diagonally, 
\begin{gather}
    \ket{\Psi} + \ket{\Phi}= \sum_{\{a, b\},\{\sigma\}}
           A'^{\sigma_1}_{\{a, b\}_1} A'^{\sigma_2}_{\{a, b\}_1\{a, b\}_2} \cdots
           A'^{\sigma_N}_{\{a, b\}_{N-1}}  \ket{ \sigma_1\sigma_2\cdots\sigma_N }   \label{eq:mps+mps} \\
\shortintertext{where}
A'^{\sigma_1} = 
\begin{bmatrix}
    A^{\sigma_1} &  B^{\sigma_1}  
\end{bmatrix}, \quad 
A'^{\sigma_i} = 
\begin{bmatrix}
    A^{\sigma_i} &  \mathbf{0}  \\
   \mathbf{0} & B^{\sigma_i} 
\end{bmatrix} \, (i=2,3 \cdots N-1),\quad 
A'^{\sigma_N} = 
\begin{bmatrix}
    A^{\sigma_N} \\
    B^{\sigma_N} 
\end{bmatrix} 
 \end{gather}
Suppose the bond dimensions of the original two MPSs  are $M_{\textrm{S}_A}$ and $M_{\textrm{S}_B}$ respectively,
then the new state $\ket{\Psi+\Phi}$ has bond dimension $M_{\textrm{S}_A}+ M_{\textrm{S}_B}$. 

In the next subsections, we present the ideas and the algorithms of three different TD-DMRG time evolution schemes adopted in our benchmark calculations afterwards: P\&C-RK4, TDVP-MU and TDVP-PS. 

\subsubsection{Schemes based on Global Propagation and Compression}
\label{sec:p&C}
The propagation and compression (P\&C) scheme is a global time evolution scheme benefiting from the modern MPS/MPO representation of DMRG wavefunctions and operators. If $\ket{\Psi(t)}$ and the time derivative $\ket{\dot{\Psi}(t)}$ are explicitly known (according to Schrödinger equation, the time derivative is  $ -i\hat H \Psi$ in atomic units), any ordinary differential equation (ODE) integrator for the initial value problem (IVP) such as the classical 4th order Runge-Kutta algorithm (RK4) we used in our former work~\cite{ren2018time}, could be applied to obtain the MPS of the next time step $\ket{\Psi(t+\dt)}$. The equations of RK4 are:
\begin{equation}
\label{eq:RK4}
\begin{aligned}
    | k_1 \rangle & = -i \hat{H}(t) | \Psi(t) \rangle   \\
    | k_2 \rangle & = -i \hat{H}(t + \dt/2) 
        (| \Psi(t) \rangle + \frac{1}{2}
        \dt | k_1 \rangle)  \\
    | k_3 \rangle & = -i \hat{H}(t + \dt/2) 
        (| \Psi(t) \rangle + \frac{1}{2} 
        \dt | k_2 \rangle)  \\
    | k_4 \rangle & = -i \hat{H}(t + \dt) 
        (| \Psi(t) \rangle +
        \dt | k_3 \rangle)  \\
    | \Psi(t+\dt) \rangle & = | \Psi(t) \rangle + \frac{1}{6} \tau(| k_1 \rangle +
    2 | k_2 \rangle + 2| k_3 \rangle + | k_4 \rangle)  
\end{aligned}
\end{equation}
Each $\ket{k_i}$ is represented by an MPS and $\hat{H}(t_i)$ is represented by an MPO. Therefore, only two types of operations including MPO$\times$MPS (Eq.~\ref{eq:mpo-apply}) and MPS$ + $MPS (Eq.~\ref{eq:mps+mps}) exist in Eq.~\ref{eq:RK4}.
After each operation, the new MPS such as $\ket{k_1}= -i \hat{H}(t) | \Psi(t) \rangle $ will have a larger virtual bond dimension, which should be compressed using algorithm based on Eq.~\ref{eq:SVD} for the further operations. It is worth mentioning that as the new MPS is not canonical, canonicalisation should be carried out before the actual compression. 

The procedure of P\&C-RK4 described above involves 
$\hat{H}\ket{\Psi}$ (``MPO $\times$ MPS'') scaling at $\order{M_\textrm{S}^2M_\textrm{O}^2d^2}$  for each site
(labeled as  ``$\hat{H}\ket{\Psi}$''), 
``MPS + MPS'' which does not require intensive computation 
and MPS compression. 
The compression consists of four kinds of operations 
(two for canonicalisation in Eq.~\ref{eq:QR} and two for compression in Eq.~\ref{eq:SVD}), 
namely QR decomposition of tensors on each site (labeled as ``QR''), 
matrix multiplication to absorb the decomposed coefficients (labeled as ``MatMul-QR'')
and the SVD counterparts (labeled as ``SVD'' and ``MatMul-SVD'').
``QR'' and ``MatMul-QR'' both scale at $\order{M_\textrm{S}^3 M_\textrm{O}^3 d}$ while
their SVD counterparts scale at $\order{\min(M_\textrm{S}^3 M_\textrm{O}^2 d, M_\textrm{S}^3 M_\textrm{O} d^2)}$ for ``SVD'' and $\order{M_\textrm{S}^3 M_\textrm{O}^2 d}$ for ``MatMul-SVD'',
because a compression reduces bond dimension from $M_\textrm{O}M_\textrm{S}$ to $M_\textrm{S}$ during the sweep.
Thus, the bottleneck of P\&C-RK4 is the canonicalisation
and the scaling of the method in a single time step is $\order{N M_\textrm{S}^3 M_\textrm{O}^3 d}$.

Compared to the local evolution scheme such as TST~\cite{feiguin2005time}, P\&C could exactly calculate the operation of high order Hamiltonian such as  $\hat{H}^n \ket{\Psi}$ and it is natural to increase the bond dimension with entanglement during the time evolution.
However, though P\&C is in principle a general evolution scheme for TD-DMRG, 
it prefers Hamiltonian whose MPO is straightforward to construct and has small bond dimension $M_\textrm{O}$, such as Frenkel-Holstein type model whose $M_\textrm{O}$ is in a linear relationship to the number of electronic states~\cite{ren2018time}. With regard to the time step size, the error of P\&C does not have a monotonic relationship with time step and there is an optimal time step depending on the system as discussed in the previous studies~\cite{ronca2017time,ren2018time}. A smaller time step will improve the accuracy of RK4 integrator but lead to more MPS compressions  and then deteriorate the whole accuracy.

\subsubsection{Schemes based on Time Dependent Variational Principle}
\label{sec:tdvp}
The Rayleigh-Ritz variational principle is widely used in finding an approximate ground state in time independent Schrödinger equation. Similarly, time dependent variational principle (TDVP) also provides a strong tool to find an optimal time dependent wavefunction if the wavefunction ansatz and the initial state are known. The Dirac-Frenkel TDVP is~\cite{dirac1930note,gatti2017applications}
\begin{gather}
    \braket{\delta \Psi | i\pdv{t} - \hat{H} | \Psi} = 0
    \label{eq:TDVP}
\end{gather}
In a geometric fashion, TDVP could be understood as an orthogonal projection of $-i\hat{H} | \Psi \rangle$ onto the tangent space of $\ket{\Psi(t)}$ at the current time:
\begin{equation}
    \pdv{\ket{\Psi}}{t} = -i \hat P \hat H \ket{\Psi}
    \label{eq:tangent}
\end{equation}
where $\hat P$ is the projector constructed by the orthonormal vectors in the tangent space. 
For a general MPS in Eq.\ref{eq:mps}, the tangent space projector could be defined as:
\begin{gather}
    \hat{P} = \sum_{i=1}^{N}  \hat{P}[1:i-1] \otimes \hat{I}_{i} \otimes \hat{P}[i+1:N]  - \sum_{i=1}^{N-1} \hat{P}[1:i] \otimes \hat{P}[i+1:N] \label{eq:proj1}
\end{gather}
Where
\begin{align}
\hat{P}[1:i] & = \sum_{a_i, a_i'} |a'_{i}[1:i]\rangle S[1:i]^{-1}_{a'_{i} a_{i}}  \langle a_{i}[1:i] | \label{eq:proj1_1} \\
\hat{P}[i+1:N] & = \sum_{a_i, a_i'} |a'_{i}[i+1:N]\rangle S[i+1:N]^{-1}_{a'_{i} a_{i}}  \langle a_{i}[i+1:N] |  \label{eq:proj1_2} \\
\hat{I}_i & = \sum_{\sigma_i} |\sigma_i \rangle \langle \sigma_i | \\
\hat{P}[1:0] & = \hat{P}[N+1:N] = 1
\end{align}

\begin{figure}
  \includegraphics[width=0.95\textwidth]{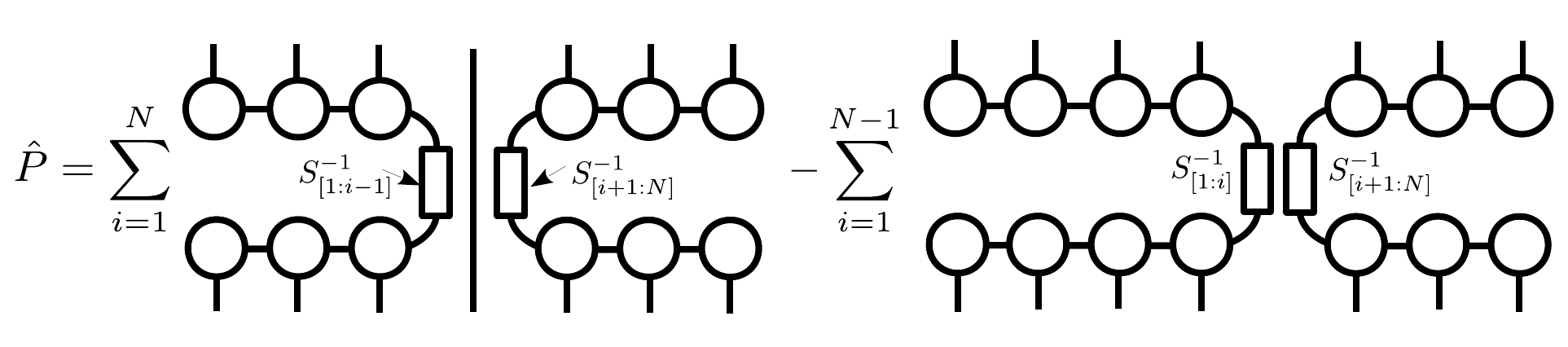}
  \caption{The graphical representation of the tangent space projector in Eq.~\ref{eq:proj1} with $N=7$.}
  \label{fig:proj}
\end{figure}

The inversion of the overlap matrix $S^{-1}$ accounts for the non-orthogonality of the renormalized basis and the ``-'' terms are to eliminate the parameterization redundancy~\cite{haegeman2011time,wouters2013thouless}. The graphical representation of the projector is shown in Fig.~\ref{fig:proj}.
In the literature, there are two different time evolution schemes based on TDVP.
They differ in choosing the specific gauge condition of the MPS
and in solving Eq.~\ref{eq:tangent}, which will be discussed in detail in the following.

\paragraph{\textbf{TDVP-MU}}

In the first TDVP evolution scheme, the gauge freedom of MPS is fixed.
For convenience, the projector in Eq.~\ref{eq:proj1} could be transformed to Eq.~\ref{eq:proj1_combine}
by combining the neighbouring ``+'' term and ``-'' term together except one ``+'' term with $i=n$:
\begin{equation}
\begin{aligned}
    \hat{P} & =  \hat{P}[1:n-1] \otimes \hat{I}_{n} \otimes \hat{P}[n+1:N]  + \sum_{i=1}^{n-1}  \hat{Q}[1:i] \otimes \hat{P}[i+1:N] + \sum_{i=n+1}^{N} \hat{P}[1:i-1] \otimes \hat{Q}[i:N]
    \label{eq:proj1_combine}
\end{aligned}
\end{equation}
where:

\begin{align}
    \hat{Q}[1:i] & = \sum_{a_{i-1}, a_{i-1}', \sigma_i,\sigma'_i} |a'_{i-1}[1:i-1] \sigma'_i \rangle \langle a_{i-1}[1:i-1] \sigma_i |   \nonumber \\
     \quad  & \cdot \big( S[1:i-1]^{-1}_{a'_{i-1} a_{i-1}} \delta_{\sigma'_i \sigma_i} - \sum_{a'_i, a_i} A^{\sigma'_i}_{a'_{i-1} a'_{i}} S[1:i]^{-1}_{a'_{i} a_{i}} A^{\sigma_i *}_{a_{i-1} a_{i}} \big)  \label{eq:proj1_Q1i}\\
     \hat{Q}[i:N] & =  \sum_{a_{i}, a_{i}', \sigma_i,\sigma'_i} |a'_{i}[i+1:N] \sigma'_i \rangle \langle a_{i}[i+1:N] \sigma_i |   \nonumber \\
     \quad  & \cdot \big( S[i+1:N]^{-1}_{a'_{i} a_{i}} \delta_{\sigma'_i \sigma_i} - \sum_{a'_{i-1}, a_{i-1}} A^{\sigma'_i}_{a'_{i-1} a'_{i}} S[i:N]^{-1}_{a'_{i-1} a_{i-1}} A^{\sigma_i *}_{a_{i-1} a_{i}} \big)  \label{eq:proj1_QiN}
\end{align}

This type of projector or the corresponding tangent space vectors were firstly proposed by Haegeman \textit{et al.}~\cite{haegeman2011time} in TD-DMRG and restated by Wouters \textit{et al.}~\cite{wouters2013thouless} to derive several post-DMRG methods for ground state and excited states.

Eq.~\ref{eq:proj1_Q1i} and ~\ref{eq:proj1_QiN} could be further simplified by adopting a specific gauge condition, and then some overlap matrices turn to identity. Assuming that the MPS is left-canonical with gauge center at site $N$, $S[1:i]$ is reduced to $I$ and it is most convenient to set $n=N$ in  Eq.~\ref{eq:proj1_combine}.
Inserting the simplified projector into Eq.~\ref{eq:tangent} yields:
\begin{equation}
\label{eq:tdvp1_C}
    i \pdv{ C^{\sigma'_N}_{l'_{N-1}} }{t}
    = \sum_{\sigma''_N, l''_{N-1}}
    H[N]_{l'_{N-1} \sigma'_N, l''_{N-1} \sigma''_N} C^{\sigma''_N}_{l''_{N-1}}
\end{equation}
\begin{equation}
\label{eq:tdvp1_L}
\begin{aligned}
    i \pdv{ L^{\sigma'_i}_{l'_{i-1}l'_i} }{t}
    = & \sum_{l_{i-1}, \sigma_i}(\delta_{l'_{i-1} l_{i-1}} \delta_{\sigma'_{i} \sigma_{i}} - p[i]_{l'_{i-1} \sigma'_i, l_{i-1} \sigma_i})
     \sum_{l_i} S[i+1:N]^{-1}_{l'_{i} l_{i}}  
     \sum_{l''_{i-1}, \sigma''_i,  l''_i} H[i]_{l_{i-1} \sigma_i l_i, l''_{i-1}  \sigma''_i l''_i}  L^{\sigma''_i}_{l''_{i-1}l''_i} 
\end{aligned}
\end{equation}
where:
\begingroup
\allowdisplaybreaks
\begin{align}
    H[i]_{l'_{i-1} \sigma'_i l'_i, l_{i-1} \sigma_i  l_i} 
    =& \sum_{ \{w\}} h[1:i-1]_{\{l',w,l\}_{i-1}}
           W^{\sigma'_{i}, \sigma_{i}}_{w_{i-1}w_{i}}
           h[i+1:N]_{\{l',w,l\}_{i}} \label{eq:tdvp1_H}\\
    h[1:i-1]_{\{l',w,l\}_{i-1}} = & \sum_{\{l'\}, \{w\}, \{l\}} h[1]_{\{l',w,l\}_1} 
                                  \cdots 
                                    h[i-1]_{\{l',w,l\}_{i-2}, \{l',w,l\}_{i-1}} \label{eq:tdvp1_h1i}\\
    h[i+1:N]_{\{l',w,l\}_{i}} = &  \sum_{\{l'\}, \{w\}, \{l\}} h[i+1]_{\{l',w,l\}_{i}, \{l',w,l\}_{i+1}}
                                 \cdots 
                                  h[N]_{\{l',w,l\}_{N-1}} \label{eq:tdvp1_hiN}\\
    h[i]_{\{l',w,l\}_{i-1}, \{l',w,l\}_i} =& \sum_{\sigma_{i}, \sigma'_{i}}  
                    A^{\sigma'_{i}*}_{l'_{i-1}l'_{i}}
                    W^{\sigma'_{i}, \sigma_{i}}_{w_{i-1}w_{i}}
                    A^{\sigma_{i}}_{l_{i-1}l_{i}} \quad (A=L \,\, \textrm{or} \,\, C)\label{eq:tdvp1_h}\\
        p[i]_{l'_{i-1} \sigma'_i, l_{i-1} \sigma_i} = & \sum_{l_i} 
                L^{\sigma'_{i}}_{l'_{i-1}l_{i}} L^{\sigma_{i}*}_{l_{i-1}l_{i}} \label{eq:tdvp1_p}
\end{align}
\endgroup

Similar equations can be derived for right/mixed-canonical MPS. 
With Eq.~\ref{eq:tdvp1_L} and Eq.~\ref{eq:tdvp1_p}, it is straightforward to prove
$\sum_{l_{i-1}, \sigma_i} L^{\sigma_i*}_{l_{i-1} l_i}  \pdv{}{t} L^{\sigma_i}_{l_{i-1} l'_i} = 0$ 
and then:
\begin{gather}
\pdv{}{t} (\sum_{l_{i-1}, \sigma_i} L^{\sigma_i*}_{l_{i-1} l_i}  L^{\sigma_i}_{l_{i-1} l'_i}) = 0  \label{eq:tdvp1_L_cano}
\end{gather}
which ensures that the left-canonical condition preserves during the time evolution formally.
Though in practical numerical calculation with a finite time step, 
the relation in Eq.~\ref{eq:tdvp1_L_cano} would not be rigorously fulfilled,
this problem is not severe with a proper time step in our experience.
Otherwise, the more general equations in Eq.~\ref{eq:proj1_combine} which have already considered the non-orthogonality of the left and right renormalized basis should be used. And it could be proved that $\pdv{}{t}S[1:i]=0\,(i=1,2,\cdots,n-1)$ and $\pdv{}{t}S[j:N]=0\,(j=n+1,n+2,\cdots,N)$.

Eq.~\ref{eq:tdvp1_C} and Eq.~\ref{eq:tdvp1_L} together form a set of coupled nonlinear equations
that are very similar to the standard EOMs of (ML-)MCTDH~\cite{meyer1990multi,beck2000multiconfiguration,wang2003multilayer}. 
To integrate these equations, we borrow ideas called variable mean field (VMF) 
and constant mean field (CMF) from the MCTDH community~\cite{beck1997efficient,beck2000multiconfiguration}.
VMF employs an all-purpose integrator
to directly solve the coupled equations.
While in CMF, it is assumed that $H[i]$ and $S[i+1:N]$ generally change much slower in time than the local matrices $C^{\sigma_N}$ and  $L^{\sigma_i}$.
As a result, during the integration of Eq.~\ref{eq:tdvp1_C} and Eq.~\ref{eq:tdvp1_L}
one may hold the ``mean field''  $H[i]$ and $S[i+1:N]$ constant for $\dt$ and evolve only the local matrix with time step smaller than $\dt$.
Hence, CMF is more efficient yet less accurate than VMF.
In this work, we use a second order CMF with a midpoint scheme in which $L^{\sigma_i}(0)$ and $C^{\sigma_N}(0)$ are integrated to $L^{\sigma_i}(\tau)$ and $C^{\sigma_N}(\tau)$
with ``mean field'' constructed by $L^{\sigma_i}(\tau/2)$ and $C^{\sigma_N}(\tau/2)$.
In VMF, adaptive Dormand-Prince's 5/4 Runge-Kutta method (RK45) is adopted to
integrate Eq.~\ref{eq:tdvp1_C} and Eq.~\ref{eq:tdvp1_L}, while in CMF, since Eq.~\ref{eq:tdvp1_C} is linear, it is integrated by Krylov subspace method instead.

Another aspect should be concerned is that the inversion of $S$ would be unstable numerically if some eigenvalues of $S$ are very small. This problem will be severe when the state is weakly correlated 
(such as a Hartree product state which is usually an initial state) and $M_\textrm{S}$ is much larger than what is required. 
To some extent, this instability problem makes this evolution scheme paradoxical in that large $M_\textrm{S}$ should in principle push the result to a numerically exact limit but in fact deteriorates it.
The same problem also arises in (ML)-MCTDH, where in order to make the EOMs more ``well-behaved'', $S$ is usually replaced with a
regularized overlap matrix $\tilde{S}$~\cite{beck2000multiconfiguration}:
\begin{equation}
\label{eq:reg1}
    \tilde{S} = S + \varepsilon e^{-S / \varepsilon}
\end{equation}
Here $\varepsilon$ is a small scalar commonly from $10^{-8}$ to $10^{-14}$.
More recently, an improved regularization scheme based on the matrix unfolding (MU) of the coefficient matrix by SVD in (ML-)MCTDH
is proposed by Meyer and Wang, which has been proved to make the time integration more accurate and robust~\cite{meyer2018regularizing,wang2018regularizing}. 
The same idea is adopted here to integrate Eq.~\ref{eq:tdvp1_L}, giving the name of the scheme ``TDVP-MU''.
When calculating the overlap matrix $S[i+1:N]^{-1}_{a'_{i} a_{i}}$, 
the gauge center is moved to the $(i+1)$th site and the matrix at this site is further decomposed by SVD:
\begin{equation}
\label{eq:tdvp1_mu_rcano}
\begin{aligned}
    \ket{l_i[i+1:N]} & = \sum_{\{l\},\{\sigma\}} L^{\sigma_{i+1}}_{l_i l_{i+1}}
    \cdots L^{\sigma_{N-1}}_{l_{N-2}l_{N-1}} C^{\sigma_{N}}_{l_{N-1}} 
    \ket{\sigma_{i+1}\cdots \sigma_N} \\
    & = \sum_{\{r\},\{\sigma\}} U_{l_i r_i} \Lambda_{r_i r_i} R^{\sigma_{i+1}}_{r_i r_{i+1}}
    \cdots R^{\sigma_{N-1}}_{r_{N-2}r_{N-1}} R^{\sigma_{N}}_{r_{N-1}} 
    \ket{\sigma_{i+1}\cdots \sigma_N} 
\end{aligned}
\end{equation}
where $|r_i|$ equals $|l_i|$. Thus, the overlap matrix $S[i+1:N]$ and its inversion could be expressed as 
$U^* \Lambda^2 U^T$ and $U^* \Lambda^{-2} U^T$ respectively. 
The Hamiltonian matrix in Eq.~\ref{eq:tdvp1_H} is also reconstructed. 
For site from $i+1$ to $N$ the matrix $A^{\sigma_i}$ in Eq.~\ref{eq:tdvp1_h} is replaced with matrix $R^{\sigma_i}$ in Eq.~\ref{eq:tdvp1_mu_rcano} and then
Eq.~\ref{eq:tdvp1_L} with matrix unfolding algorithm becomes:
\begin{equation}
\label{eq:tdvp1_mu_L}
\begin{aligned}
    i \pdv{ L^{\sigma'_i}_{l'_{i-1}l'_i} }{t}
    = & \sum_{l_{i-1}, \sigma_i}(\delta_{l'_{i-1} l_{i-1}} \delta_{\sigma'_{i} \sigma_{i}} - p[i]_{l'_{i-1} \sigma'_i, l_{i-1} \sigma_i})
     \sum_{r_i, r'_i, l_i} [  U_{l'_i r'_i}^* \Lambda_{r'_i r'_i}^{-1} 
    \underline{ \underline{\Lambda_{r'_i r'_i}^{-1} U_{r'_i l_i}^T ] U^*_{l_i r_i} \Lambda_{r_i r_i}} }\\
    & \sum_{l''_{i-1}, \sigma''_i,  l''_i, r''_i} 
    H[i]_{l_{i-1} \sigma_i r_i, l''_{i-1}  \sigma''_i  r''_i}  \Lambda_{r''_i r''_i} U^T_{r''_i l''_i}
    L^{\sigma''_i}_{l''_{i-1}l''_i} 
\end{aligned}    
\end{equation}
The expression inside ``$[\cdots]$'' is $S[i+1:N]^{-1}$. The key point of this new regularization scheme is that the underlined part could be contracted first, which is $\delta_{r'_i r_i}$. Thus, only the singular matrix $\Lambda_{r'_i r'_i}$ instead of $\Lambda_{r'_i r'_i}^{2}$ should be regularized:
\begin{equation}
\label{eq:reg2}
\tilde{\Lambda}_{r'_i r'_i} = \Lambda_{r'_i r'_i} + \varepsilon^{1/2} e^{-\Lambda_{r'_i r'_i} / \varepsilon^{1/2}} 
\end{equation}
The power $1/2$ here is for consistency with the original regularization scheme (Eq.~\ref{eq:reg1})
and $\Lambda_{r''_i r''_i}$ in Eq.~\ref{eq:tdvp1_mu_L} is untouched in order to be minimally invasive
as stated in Ref.~\onlinecite{meyer2018regularizing,wang2018regularizing}.
In this work we choose $\varepsilon=10^{-10}$ unless otherwise stated.

The complexity analysis of the TDVP-MU scheme with VMF and CMF is as follows.
Note that in TD-DMRG it is necessary to contract
small matrices one by one instead of
explicitly constructing the large tensors~\cite{chan2004algorithm,schollwock2011density}.
VMF and CMF both require the calculation of environment matrices $h[1:i-1]$ and 
$h[i+1:N]$ scaling at
$\order{M_\textrm{S}^2 M_\textrm{O}^2 d^2 + M_\textrm{S}^3 M_\textrm{O} d}$ (labeled as ``Get Env''),
SVD decomposition of the gauge center with the subsequent
matrix multiplication scaling at $\order{M_\textrm{S}^3 d}$ in Eq.~\ref{eq:tdvp1_mu_rcano} (labeled as ``SVD'' and ``MatMul-SVD''),
and the calculation of time derivative in 
Eq.~\ref{eq:tdvp1_C} and Eq.~\ref{eq:tdvp1_mu_L} (labeled as ``Deriv'').
Evaluation of the time derivatives contains a number of matrix multiplications and the overall scaling is
$\order{M_\textrm{S}^2 M_\textrm{O}^2 d^2 + M_\textrm{S}^3 M_\textrm{O} d + M_\textrm{S}^3 d^2}$,
where the first two terms and the last term correspond to the contraction of 
$H[i]$ and $(I-p[i])$ respectively.
Thus the  total scaling in VMF is
$\order{N (M_\textrm{S}^2 M_\textrm{O}^2 d^2 + M_\textrm{S}^3 M_\textrm{O} d + M_\textrm{S}^3 d^2)}$,
while in CMF the total scaling for a single step is
$\order{N f (M_\textrm{S}^2 M_\textrm{O}^2 d^2 + M_\textrm{S}^3 M_\textrm{O} d + M_\textrm{S}^3 d^2)}$
where $f$ is the number of steps required for the propagation of each local site within $\tau$.
Although at first glance CMF is more time-consuming than VMF,  
much larger step size is possible for CMF  which reduces the time cost spent on decomposing the matrix and constructing the environment, compensating the effect of the factor $f$.
This time saving in CMF is important in MCTDH in that constructing the environment
involves a contraction of the high order coefficient matrix ~\cite{beck1997efficient,beck2000multiconfiguration,gatti2017applications}.
However, in MPS context such advantage is not prominent because all matrices including the coefficient matrix $C^{\sigma_N}$
are of the same size. As a result, constructing the environment has a similar complexity as calculating the derivatives.

\paragraph{\textbf{TDVP-PS}}
The second evolution scheme based on TDVP is called projector-splitting (PS).
The idea of PS is that the tangent space projector in Eq.~\ref{eq:proj1} is invariant under different gauge conditions. More specifically, after canonicalisation of a general MPS in Eq.~\ref{eq:mps} from site $N$ to $i+1$, $|r_{i}[i+1:N] \rangle$ becomes the right-hand orthonormal renormalized basis, which
is related to $|a_{i}[i+1:N] \rangle$ by:

\begin{gather}
 |a_{i}[i+1:N] \rangle = \sum_{r_i} D_{a_i r_i} |r_{i}[i+1:N] \rangle
\end{gather}
The matrix $D$ is an upper triangular matrix in RQ decomposition after the canonicalisation described in Eq.~\ref{eq:QR}. Therefore, the overlap matrix $S[i+1:N]$ equals $D^* D^T$ and the projector $\hat{P}[i+1:N]$ in Eq.\ref{eq:proj1_2} defined for a general non-canonical MPS is transformed to:
\begin{equation}
\begin{aligned}
    \hat{P}[i+1:N]  
    & =  \sum_{r'_i, r_i}  |r'_{i}[i+1:N] \rangle  \left[\sum_{a'_i, a_i}  D^T_{r'_i a'_i} (D^*D^T)^{-1}_{a'_i a_i} D^*_{a_i r_i} \right]_{=\delta_{r'_ir_i}}  \langle r_{i}[i+1:N]  |  \\
     & = \sum_{r_i}  |r_{i}[i+1:N] \rangle \langle r_{i}[i+1:N]  |
\end{aligned}
\end{equation}
Similar result can be obtained for $\hat{P}[1:i]$:
\begin{gather}
     \hat{P}[1:i] = \sum_{l_i}  |l_{i}[1:i] \rangle \langle l_{i}[1:i]  |
\end{gather}

On the one hand, this definition of the tangent space projector does not contain
any inversion operations of the overlap matrix, 
which seems to be a remarkable improvement over the first definition in  Eq.~\ref{eq:proj1_1} and Eq.~\ref{eq:proj1_2}.
On the other hand, since the gauge is not fixed in different terms of this projector, 
the integration algorithm described in section \textbf{\textit{TDVP-MU}} could not be directly applied.
Lubich and Haegeman \textit{et al.} proposed to use a symmetric second order Trotter decomposition to split the formal propagator into the individual terms~\cite{lubich2015tt, haegeman2016unifying}:
\begin{equation}
\label{eq:ps_trotter}
\begin{aligned} 
e^{-i\hat P \hat{H} \dt} = 
\left[\prod_{i=1}^{N-1} e^{-i\hat{P}[1:i-1]\otimes \hat{I}_i \otimes \hat{P}[i+1:N] \hat{H} \dt/2} 
\cdot e^{i\hat{P}[1:i]\otimes \hat{P}[i+1:N] \hat{H} \dt/2} \right] 
\cdot   e^{-i\hat{P}[1:N-1] \otimes \hat{I}_N \hat{H} \dt} \\ 
\cdot \left[\prod_{i=N-1}^{1}  e^{i\hat{P}[1:i]\otimes \hat{P}[i+1:N] \hat{H} \dt/2} 
\cdot e^{-i\hat{P}[1:i-1]\otimes \hat{I}_i \otimes \hat{P}[i+1:N] \hat{H} \dt/2} \right]
+ \order{\dt^3}
\end{aligned}
\end{equation}
Based on the propagator in Eq.~\ref{eq:ps_trotter}, a single step of time evolution consists of a left-to-right sweep 
and a subsequent right-to-left sweep each with step size $\dt/2$.
Taking left-to-right sweep as an example, the matrix at the gauge center $C_{l_{i-1}r_i}^{\sigma_i}$
is firstly evolved forward in time by applying the projector $\hat{P}[1:i-1]\otimes \hat{I}_i \otimes \hat{P}[i+1:N]$:
\begin{equation}
\label{eq:tdvp2_C}
    i\pdv{C_{l'_{i-1}r'_i}^{\sigma'_i}}{t} = 
    \sum_{ l_{i-1}, \sigma_i, r_i}
    H[i]_{l'_{i-1} \sigma'_i r'_i, l_{i-1} \sigma_i r_i}
    C_{l_{i-1}r_i}^{\sigma_i}
\end{equation}
where $H[i]$ and the ingredients $h[1:i-1]$, $h[i+1:N]$, $h[i]$ all have the same definitions as in Eq.~\ref{eq:tdvp1_H} \ref{eq:tdvp1_h1i},~\ref{eq:tdvp1_hiN},~\ref{eq:tdvp1_h} except that the $A^{\sigma_i}$ in Eq.~\ref{eq:tdvp1_h} is replaced with $L^{\sigma_i}$ or $ R^{\sigma_i}$ accordingly. Then,
the evolved matrix $C_{l_{i-1}r_i}^{\sigma_i}$
is decomposed by QR as Eq.~\ref{eq:QR} to obtain the left-canonical matrix $L^{\sigma_i}_{l_{i-1} l_i}$ and the coefficient matrix $D_{l_i r_{i}}$. $D_{l_i r_{i}}$ is evolved backward in time by applying the projector $\hat{P}[1:i]\otimes \hat{P}[i+1:N]$:
\begin{equation}
\label{eq:tdvp2_D}
    i\pdv{D_{l'_i r'_i}}{t} = 
    \sum_{l_i, w_i, r_i}
    h[1:i]_{\{l', w, l \}_i}
    h[i+1:N]_{\{r', w, r \}_i}
    D_{l_i r_i} 
\end{equation}
Afterwards, the gauge center is moved to site $i+1$ by contracting the evolved $D_{l_i r_i}$ and $R^{\sigma_{i+1}}_{r_i r_{i+1}}$ together to obtain $C_{l_{i}r_{i+1}}^{\sigma_{i+1}} = \sum_{r_i} D_{l_i r_i} R^{\sigma_{i+1}}_{r_i r_{i+1}}$. Following the procedure above, the sweep continues until all the individual projectors in Eq.~\ref{eq:ps_trotter} are applied. Inspired by the original two-site DMRG algorithm, it is also possible to formulate TDVP-PS into a two-site algorithm so that the bond dimension could grow up adaptively~\cite{haegeman2016unifying}. However, considering the two-site algorithm is much more expensive than the one-site algorithm both in the tensor contraction and QR decomposition, we use the one-site algorithm described above in this paper.

In principle, solving Eq.~\ref{eq:tdvp2_C} and Eq.~\ref{eq:tdvp2_D} could be accomplished 
by any ODE integrator such as the RK45 algorithm we use in TDVP-MU.
However, since they are linear equations, the Krylov subspace method (Lanczos algorithm for Hermitian operator here) is preferred as it is unitary and is considered to be better than the explicit time-stepping integrators for matrix exponential operator~\cite{hochbruck1997krylov,haegeman2016unifying}.
In our calculation, the dimension of the Krylov subspace is adaptive and 
the Lanczos iteration continues until $\ket{\Psi(t+\dt)}$ converges.

TDVP-PS  has a similar computational complexity as  TDVP-MU (CMF) except for 
the calculation of derivatives.
With simpler EOMs in Eq.~\ref{eq:tdvp2_C} and Eq.~\ref{eq:tdvp2_D}, the calculation of the derivative for forward evolution scales at
$\order{M_\textrm{S}^2 M_\textrm{O}^2 d^2 + M_\textrm{S}^3 M_\textrm{O} d}$
(labeled as ``Deriv-Forward'')
and for backward evolution it scales at $\order{M_\textrm{S}^3 M_\textrm{O}}$ (labeled as ``Deriv-Backward'').
So the total scaling for a single step is
$\order{N f (M_\textrm{S}^2 M_\textrm{O}^2 d^2 + M_\textrm{S}^3 M_\textrm{O} d)}$,
where $f$ represents the number of Krylov subspace vectors. Besides calculating the derivatives, Krylov subspace method also includes other matrix operations of small size such as calculating the $\{\alpha\}$ and $\{\beta\}$ matrix elements with the Lanczos three-term recurrence relation~\cite{arnoldi1951principle} and diagonalizing the tridiagonal Hamiltonian in the subspace (labeled as ``Krylov'').
Other labels for TDVP-PS is the same with that of TDVP-MU.

In summary, since the TDVP-MU scheme and the TDVP-PS scheme are both based on the time dependent variational principle, $\ket{\Psi(t)}$ should be the same if not considering the numerical error. 
Additionally, the two schemes both require to define a fixed $M_\textrm{S}$ a priori,
and additional renormalized basis should be constructed smartly to complement the empty MPS space if the initial state is weakly correlated.  Numerically, once $M_\textrm{S}$ is fixed, the error of these two TDVP-based schemes has a monotonic relationship with the time step size. The main difference lies in that  TDVP-MU would introduce a minor artificial regularization, while TDVP-PS is inherently free of it. We also notice that TDVP-PS has already been implemented in MCTDH recently~\cite{lubich2015mctdh,kloss2017implementation,bonfanti2018tangent}.

\subsection{CPU-GPU Heterogeneous Computing}

In the TD-DMRG algorithms described above, the hotspots are usually the  tensor contractions and the matrix decompositions.
GPUs are well suitable for the former not the latter. Therefore, we adopt a CPU-GPU heterogeneous computing strategy.
We store the matrices in the GPU memory instead of the host memory,
and in most cases call cuBLAS to manipulate them. When doing matrix decomposition, we first transfer the matrix from the GPU memory to the host memory and then use a single CPU
core to complete the decomposition, and in the end copy the matrix back.
The overhead caused by data transfer is not negligible which we believe could be avoided by 
a smarter implementation because GPUs are able to run computation and data transfer at the same time.

For comparison, multi-core CPU calculation is also benchmarked. Here, we only consider the simplest strategy of CPU parallelism, which is breaking up the dense matrix computations into subblocks performed automatically 
by the standard linear algebra libraries such as Intel® Math Kernel Library in our case.

\subsection{Computational Details of the FMO Model}
\begin{figure}
  \includegraphics[width=.48\textwidth]{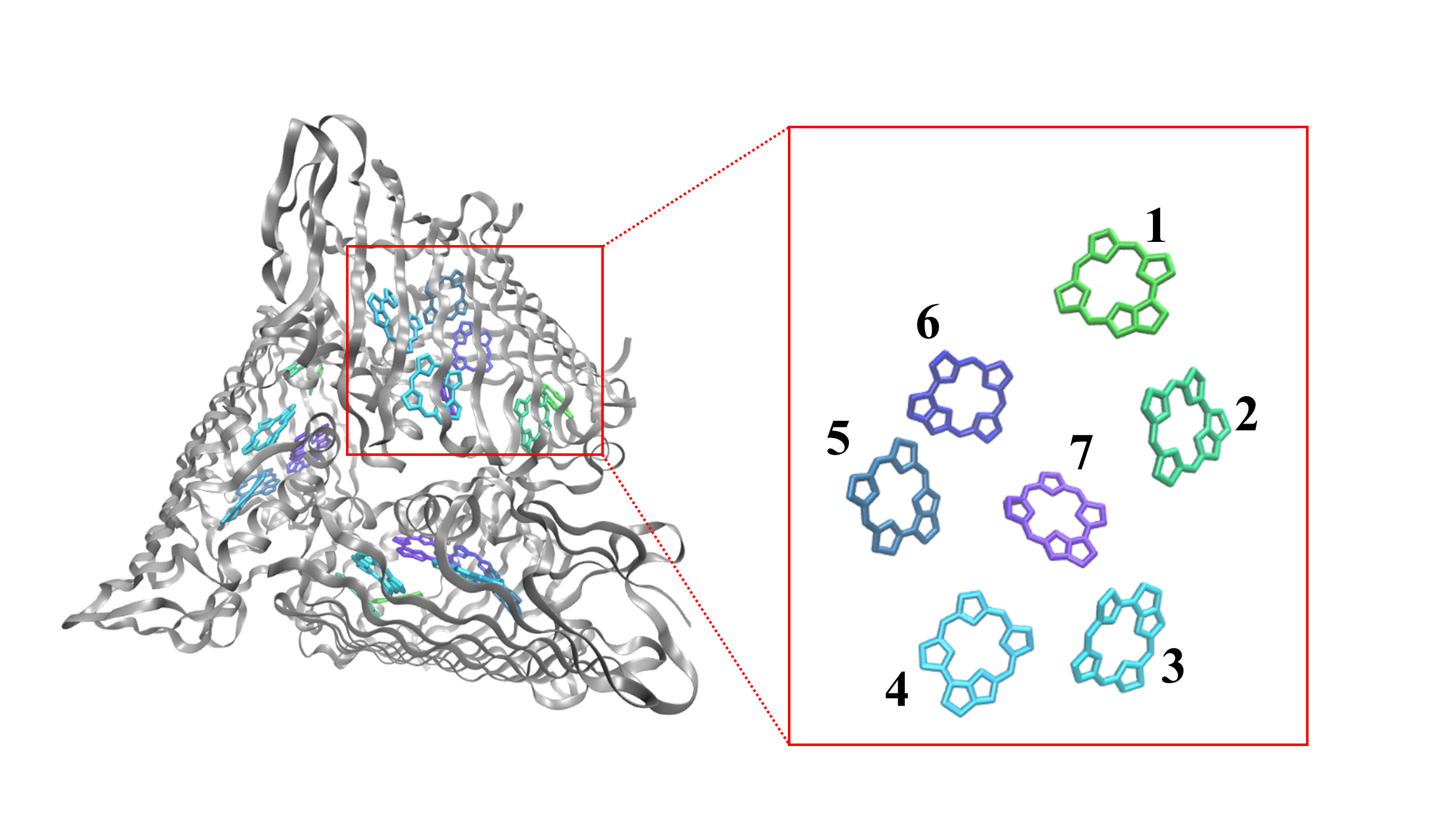}
  \caption{A diagram of the 7-site FMO model in our calculations.}
  \label{fig:diagram}
\end{figure}

The widely used Frenkel-Holstein Hamiltonian~\cite{holstein1959studies, schroter2015exciton} to describe the 7-site FMO model is that:
\begin{equation}
\label{eq:holstein}
\begin{aligned}
\hat H= &\sum_{m}E_m a_m^\dag a_m+ \sum_{m\neq n} J_{mn}a_m^\dag a_n+
\sum_{m\lambda}\omega_{m\lambda}\left(b_{m\lambda}^\dag b_{m\lambda}+\frac{1}{2} \right )
\\ & + \sum_{m\lambda}g_{m\lambda}\omega_{m\lambda}\left(b_{m\lambda}^\dag + b_{m\lambda} \right )a_m^\dag a_m
\end{aligned}
\end{equation}
where $a_m^\dag(a_m)$ is the exciton creation (annihilation) operator on the $m$th site whose local 
excitation energy is $E_m$, 
$J_{mn}$ is the Coulomb interaction between the $m$th and $n$th site,
$b_{m\lambda}^\dag(b_{m\lambda})$ is the phonon creation (annihilation) operator of
vibration mode $\lambda$ of $m$th site with vibration frequency $\omega_{m\lambda}$ and dimensionless electron-phonon coupling strength $g_{m \lambda}$. 
Fig.~\ref{fig:diagram} shows a diagram of the model.

We obtain $E_m$ and $J_{mn}$ from previous literature~\cite{moix2011efficient,schulze2016multi} and each site has the same environment with $\omega_{m \lambda}$ and $g_{m \lambda }$ calculated by equally spaced discretization of the bath spectral density~\cite{de2015discretize} from experiments~\cite{wendling2000electron,schulze2016multi}.
For each site, 35 vibration modes are discretized, giving 252 
DOFs in total (245 vibrational DOFs + 7 electronic DOFs).
The dimension of basis for each mode varies with frequency
and most modes have 8 or 4 phonon occupation levels.
To minimize the entanglement, the electronic sites in the MPS chain are arranged in the order of [7, 5, 3, 1, 2, 4, 6]. All the local vibrational DOFs are arranged next to the local electronic site as our previous work~\cite{ren2018time}.

We simulate the exciton dynamics at zero temperature. At $t=0$, an excitation at electronic site 1 is prepared with all vibrations at their lowest energy level, which composes a Hartree product state. Therefore, for simulation with $M_\textrm{S}>1$, TDVP-based methods (TDVP-MU and TDVP-PS) should somehow find $M_\textrm{S}-1$ renormalized basis
to complement the empty matrix elements in the initial MPS.
To solve this, in all the simulations, we use P\&C-RK4 to propagate the state  with $M_\textrm{max}=256$ and time step size $\tau=80 \, \textrm{a.u.}$ in the first 10 steps to $t=800 \, \textrm{a.u.}$. Thereafter, the state is compressed to the pre-defined bond dimension $M_\textrm{S}$ and then evolves with different evolution schemes.
Throughout this paper we use atomic unit (a.u.) as the unit of time unless otherwise stated and omit the unit of $t$ and $\dt$ for simplicity.

\section{Results and Discussions}

\subsection{Accuracy and Time Step Size}
\label{sec:accuracy}
\begin{figure}
	\includegraphics[width=.48\textwidth]{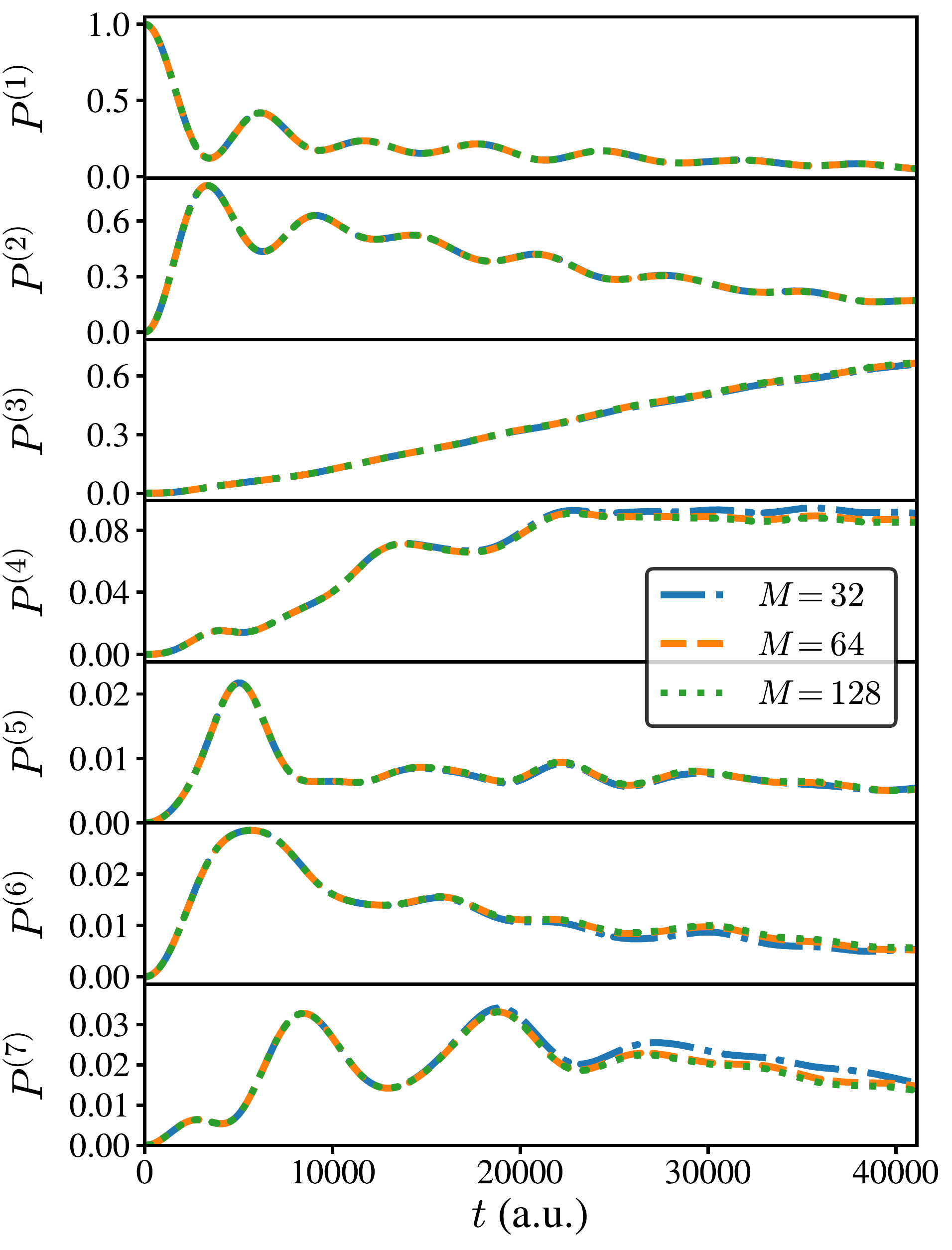}
	\caption{The exciton population dynamics ($P^{(n)}$ v.s $t$) obtained from $M_\textrm{S}=32$, $M_\textrm{S}=64$ and $M_\textrm{S}=128$ by  TDVP-PS.
	The curves have already converged on step size $\dt$.}
	\label{fig:convergence}
\end{figure}

To quantitatively evaluate the relative accuracy in our calculations,
the mean cumulative deviation of exciton populations at time $t$ is used:
\begin{equation}
\label{eq:error}
    \textrm{error}(t) = \frac{\sum_{n=1}^7\int_0^t | P^{(n)}(t') - P^{(n)}_{\textrm{ref}}(t') | dt'}{7t}
\end{equation}
where $P^{(n)}$ represents the exciton population at the $n$th site of the 7-site FMO model. $P^{(n)}_{\textrm{ref}}$ is the reference result obtained from TD-DMRG calculation with an appropriate 
MPS bond dimension $M_\textrm{S}$ and evolution time step size $\dt$.
The standard for choosing $M_\textrm{S}$ and $\dt$ is
that the exciton populations have converged on both of them.
Such convergence on $M_\textrm{S}$ for the TDVP-PS method is illustrated in Fig.~\ref{fig:convergence}
where all curves have already converged on $\dt$.
The population obtained with $M_\textrm{S}=32$ has already been semi-quantitatively correct although the deviation is visible after $t>20000$. The curves for $M_\textrm{S}=64$ and $M_\textrm{S}=128$ almost overlap with each other.
To be more discreet, we choose $M_\textrm{S}=256$ with $\dt=10$ as the reference for better accuracy.
A more comprehensive and quantitative demonstration for the validity of our reference is included in Appendix~\ref{sec:app-valid}.
Note that using reference calculated by TDVP-MU (VMF)  with sufficiently small
regularization parameter rather than 
 TDVP-PS  has little effect on results presented in this paper.

\begin{figure}
  \includegraphics[width=.48\textwidth]{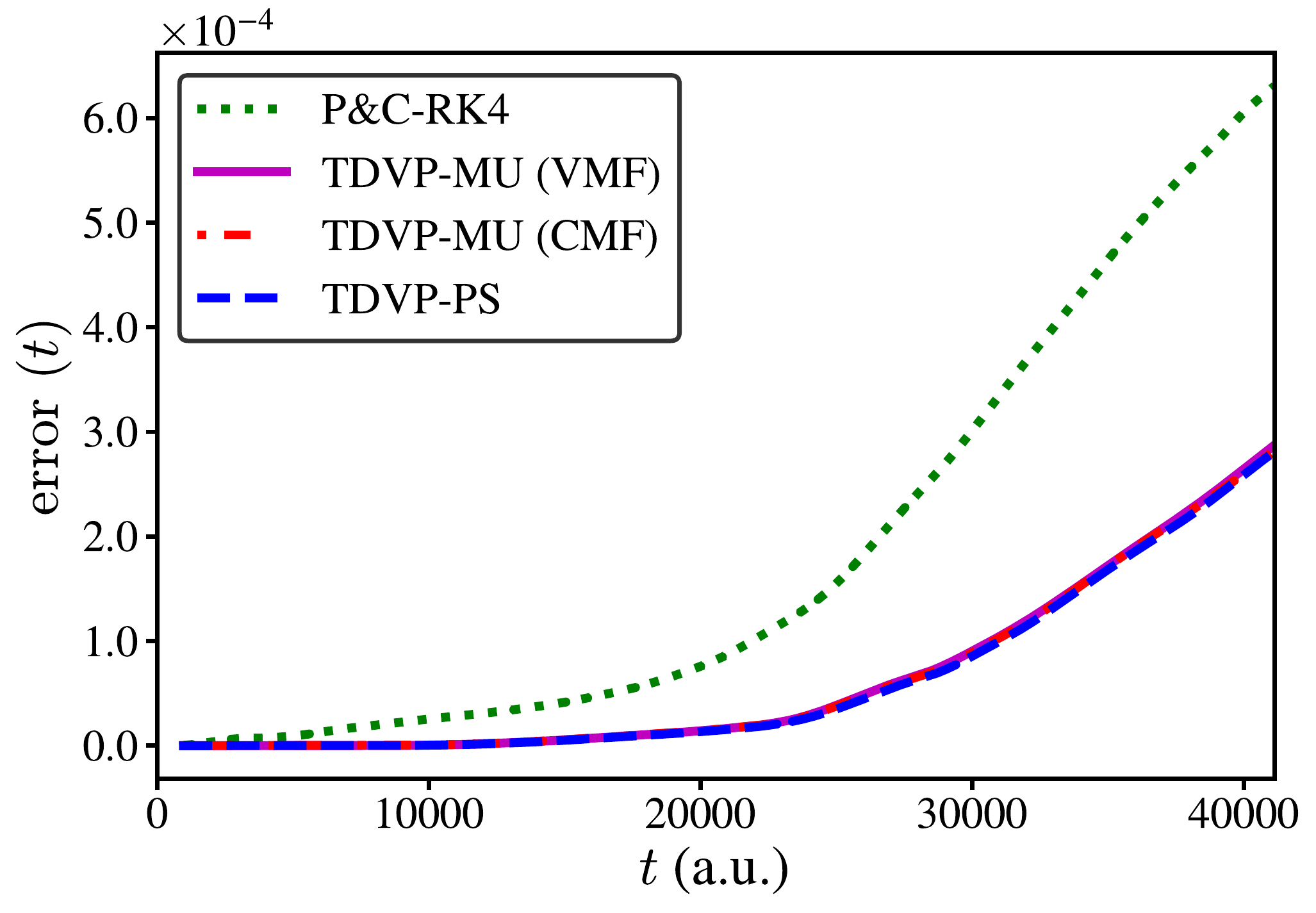}
  \caption{Mean cumulative deviation as a function of evolution time $t$ 
          of  P\&C-RK4 ,
           TDVP-MU  with VMF or CMF and  TDVP-PS 
          to $t=41120$.
          The bond dimensions  $M_\textrm{S}$ are set to 64 and the step sizes are chosen
          to minimize the error for each scheme.}
  \label{fig:error-long-t}
\end{figure}

The mean cumulative deviation of the three schemes as a function of $t$
is plotted in Fig.~\ref{fig:error-long-t} with $M_\textrm{S}=64$.
For  P\&C-RK4 the time step size is 160 which strikes a balance between the RK4 integrator error and the MPS wavefunction compression error.
The step size in TDVP-based schemes are chosen such that the error due to the integrator is negligible compared to the restricted $M_\textrm{S}$. For  TDVP-MU (VMF)  the step size controlled by the adaptive RK45 integrator (relative tolerance and absolute tolerance are $10^{-5}$ and $10^{-8}$ respectively) spans from 5 to 30.
For  TDVP-MU (CMF) and TDVP-PS  we choose $\dt=2$ which is a relatively small value
to reduce the error introduced by the CMF approximation or Trotter decomposition.
We find that at long time scale the TDVP based methods exhibit almost the same error, as they share the same starting point during their derivation,
while  P\&C-RK4  guided by a different philosophy shows larger error than TDVP based methods.

\begin{figure}
  \includegraphics[width=.48\textwidth]{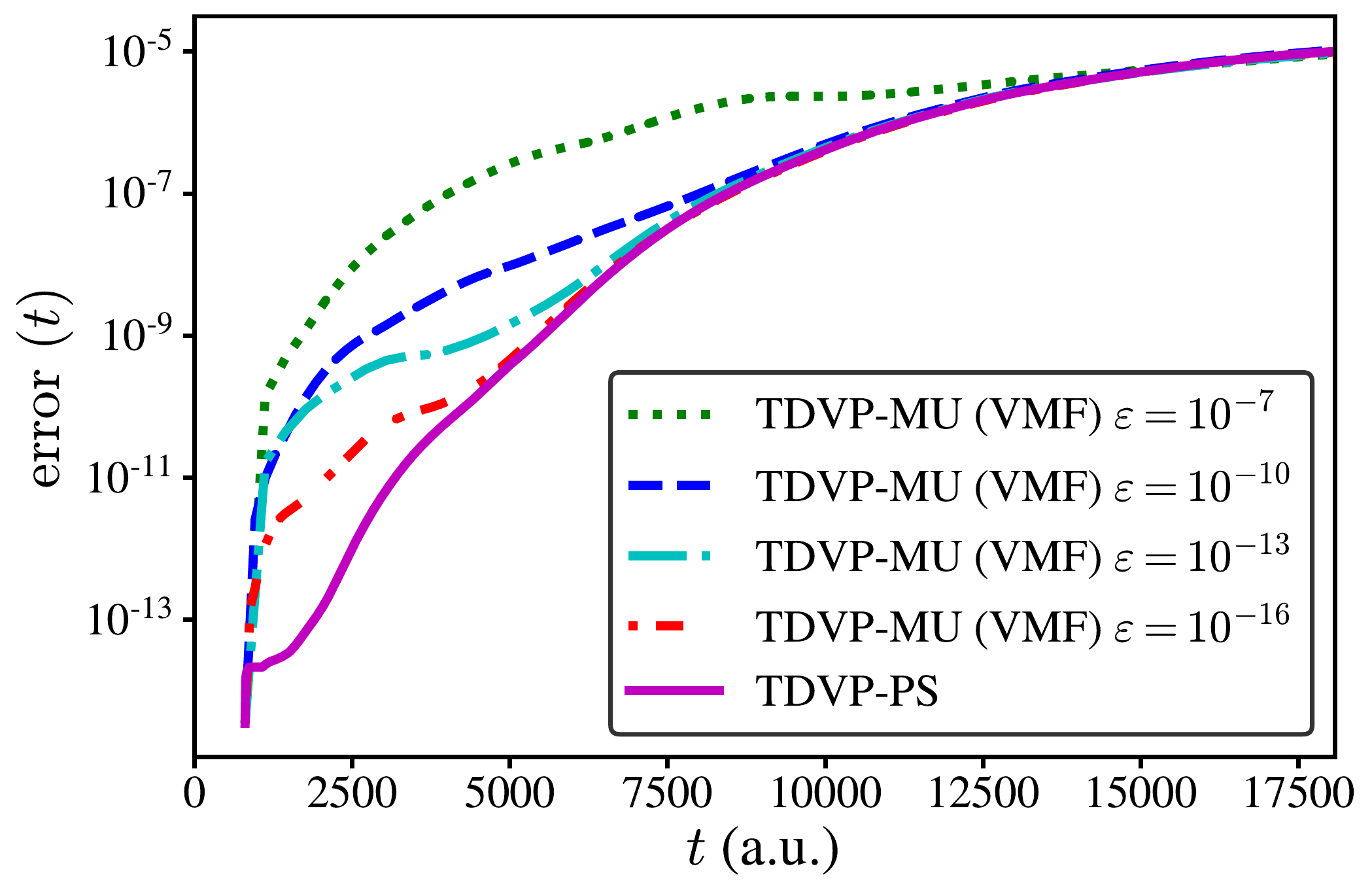}
  \caption{Mean cumulative deviation as a function of evolution time $t$ 
          of  TDVP-MU (VMF)  and  TDVP-PS 
          to $t=17500$. The $y$ axis is in logarithmic scale.
          For  TDVP-MU (VMF) , results with regularization parameter $\varepsilon$ 
          ranging from $10^{-7}$ to $10^{-16}$ are shown.
          The bond dimensions $M_\textrm{S}$ of the two schemes are set to 64 and the step sizes are chosen
          to minimize the error for each scheme.}
  \label{fig:error-short-t}
\end{figure}

Although  TDVP-MU  and  TDVP-PS  are similar in long time limit,
subtle differences between the two schemes when $t < 15000$ are observed,
which is illustrated in Fig.~\ref{fig:error-short-t}.
TDVP-MU (CMF) is not shown in Fig~\ref{fig:error-short-t}
as it produces the same result with the VMF approach at small step size limit.
Fig.~\ref{fig:error-short-t} shows that  TDVP-MU (VMF)  is not as accurate as  TDVP-PS 
at short time scale,
which is probably caused by the artificial regularization in Eq.~\ref{eq:reg2}.
Using  TDVP-MU (VMF)  with sufficiently small regularization parameter $\varepsilon$ as the reference will give the same tendency shown here.
When the regularization parameter $\varepsilon$ is decreased from  $10^{-7}$ to $10^{-16}$, which reduces the invasion to the original EOMs,
the accuracy of  TDVP-MU (VMF)  is improved but a smaller time step size is required.
 However, 
it should be noted that the error caused by the regularization at the short time regime is several orders of magnitude smaller than the error at the long time regime controlled by the MPS approximation of the wavefunction with a restricted $M_{\textrm{S}}$ as illustrated in Fig~\ref{fig:error-long-t}.  Thus in this case setting $\varepsilon$ to be $10^{-10}$ is more reasonable 
in terms of practical computation because
this setup costs less simulation time than setting $\varepsilon$ to be $10^{-13}$ or $10^{-16}$ 
without significant precision deterioration.
We also compare the new MU regularization scheme with the original regularization scheme and find that 
the new MU scheme is more accurate with the same $\varepsilon$ as the former studies in (ML-)MCTDH~\cite{meyer2018regularizing,wang2018regularizing}.

\begin{figure}
  \includegraphics[width=.48\textwidth]{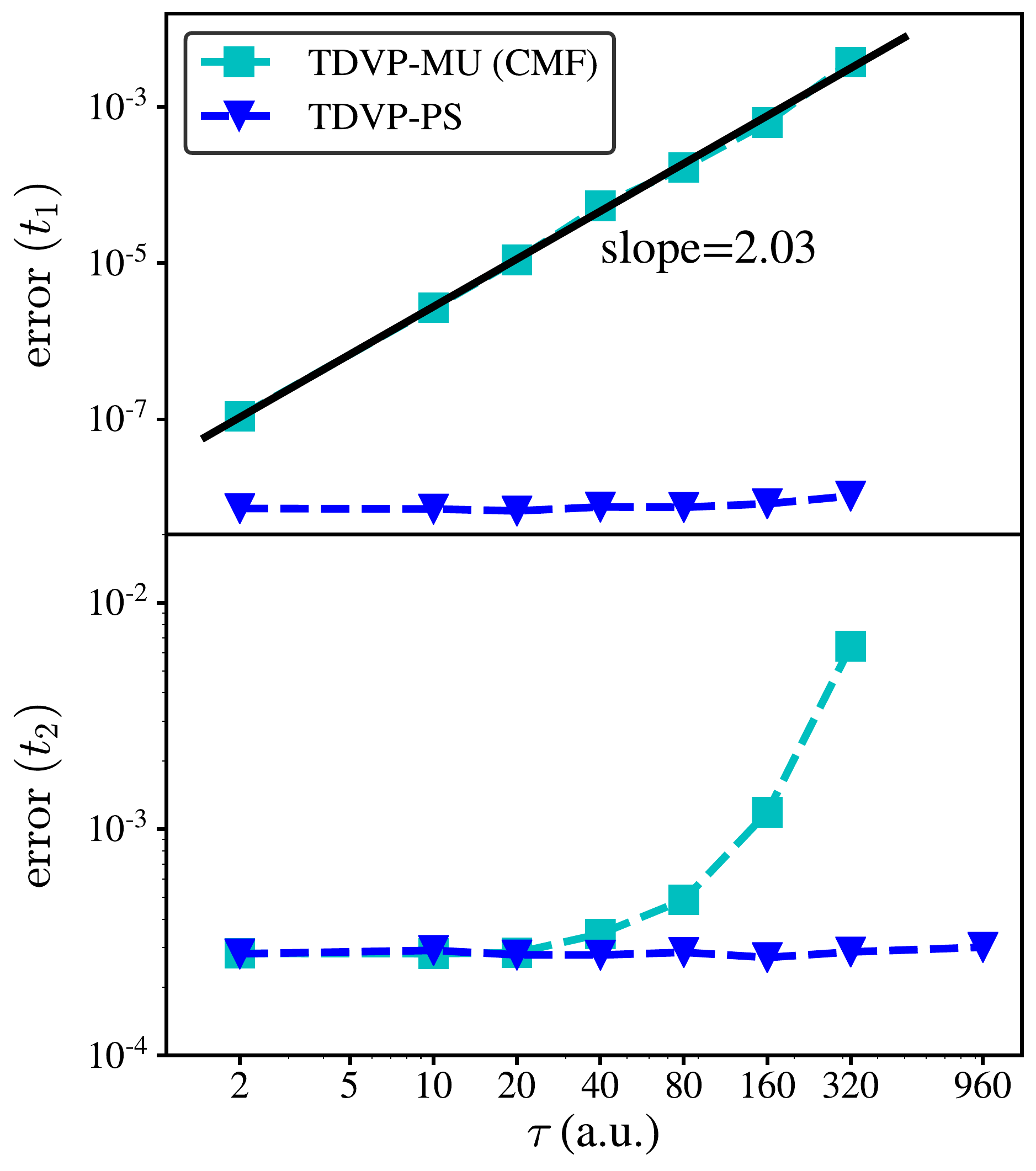}
  \caption{Mean cumulative deviation evaluated at $t_1=6560$ (upper panel) and $t_2=41120$ (lower panel)
           for  second order evolution schemes TDVP-MU (CMF)  and  TDVP-PS  with various step sizes $\dt$.
           The bond dimensions of the two schemes are set to 64. In the upper panel the relation between
           mean cumulative deviation of TDVP-MU (CMF) and $\dt$ is linear in logarithmic scale and the slope of linear fitting is 2.03.}
  \label{fig:error-dt}
\end{figure}

Despite the fact that TDVP based methods will obtain the same results in the small step size $\dt$ limit and small regularization parameter $\varepsilon$ limit, 
they behave differently to the adjustment of $\dt$ because their EOMs and integration algorithms differ from each other. 
Since TDVP-MU (CMF) and  TDVP-PS are both second order methods, we investigate how large the time step $\dt$ can be used in these two schemes with $M_\textrm{S}=64$.
The upper panel of Fig.~\ref{fig:error-dt} illustrates the error at $t_1=6560$ which is at the
early stage of the evolution. 
The error of TDVP-MU (CMF) shows a linear relationship with $\tau$ under logarithmic scale and its slope of linear fitting is 2.03,
which is in accordance with the second order approximation of  TDVP-MU (CMF) . It also indicates that for CMF the integrator error is dominant in this time scale.
To the contrary, the error of TDVP-PS  is insensitive to $\dt$ within the range of $\dt$ shown in Fig.~\ref{fig:error-dt}, which means the error due to second order Trotter decomposition is smaller than the error due to MPS ansatz with $M_\textrm{S}=64$.
In the lower panel, the time to measure the error is at $t_2 = 41120$ 
and we observe generally the same trend as in the upper panel,
except that when $\dt \leq 20$
the error of TDVP-MU (CMF) and TDVP-PS are close to each other, indicating that the time step has already converged at this time point and the error is controlled by the MPS ansatz. However, as the upper panel, TDVP-PS allows a time step at least 16 times larger  than TDVP-MU (CMF) for converged result. Therefore, though both TDVP-PS and TDVP-MU (CMF) implemented here are second order methods, the prefactor of the error term of TDVP-PS is much smaller than that of TDVP-MU (CMF).

\subsection{GPU Acceleration}
\label{sec:efficiency}
According to the analysis of computational complexity of the intensive steps in the three time evolution schemes in Section ~\ref{sec:methods} (a summary is shown in Table~\ref{tab:scaling}), we explore how these steps can be accelerated by multi-core CPU 
and particularly CPU-GPU heterogeneous computing.
Although the time cost of each algorithm should vary with different implementations
and there certainly are rooms for optimizations in our codes, 
we believe the data presented in this section reflects the correct tendency.
Our benchmark platforms are Intel® Xeon® CPU E5-2680 v4 @ 2.40GHz for CPU-only calculations and 
Intel® Xeon® Gold 5115 CPU @ 2.40GHz with NVIDIA® Tesla® V100-PCIE-32GB for CPU-GPU heterogeneous calculations.  Only one GPU is used in the benchmark.

\renewcommand{\arraystretch}{1.3}
\newcommand{\getlr}{M_\textrm{S}^2 M_\textrm{O}^2 d^2 + M_\textrm{S}^3 M_\textrm{O} d}

\begin{table*}
\caption{\label{tab:scaling}
Summary of the computational complexity of P\&C-RK4, TDVP-MU and TDVP-PS time evolution schemes.}
\begin{ruledtabular}
\begin{tabular}{cccc}
        Sub-step \footnotemark[1] \    & P\&C-RK4                           & TDVP-MU                          & TDVP-PS   \\
\hline
$\hat{H}\ket{\Psi}$       & $M_\textrm{S}^2M_\textrm{O}^2d^2$  & ---                              & ---                   \\
QR          & $M_\textrm{S}^3 M_\textrm{O}^3 d$  & ---                               & $M_\textrm{S}^3 d$   \\
MatMul-QR   & $M_\textrm{S}^3 M_\textrm{O}^3 d$  & ---                              & $M_\textrm{S}^3 d$   \\
SVD         & $\min(M_\textrm{S}^3 M_\textrm{O}^2 d, M_\textrm{S}^3 M_\textrm{O} d^2)$  & $M_\textrm{S}^3 d$                & ---                     \\
MatMul-SVD  & $M_\textrm{S}^3 M_\textrm{O}^2 d$  & $M_\textrm{S}^3 d$               & ---   \\
Get Env     & ---                                & $\getlr$                        & $\getlr$ \\
Deriv       & ---                                & $\getlr + M_\textrm{S}^3 d^2$   & \makecell{$\getlr$ (forward) \\ $M_\textrm{S}^3 M_\textrm{O}$  (backward)} \\
Overall     &  $N M_\textrm{S}^3 M_\textrm{O}^3 d$ & \makecell{ $N(\getlr + M_\textrm{S}^3 d^2)$ (VMF) \\ $Nf(\getlr + M_\textrm{S}^3 d^2)$ (CMF) } & $N f (\getlr) $ 
\end{tabular}
\end{ruledtabular}
\footnotetext[1]{See the text in Section~\ref{sec:methods} for the definitions and labels of the sub-steps. }
\end{table*}

\begin{figure}
  \includegraphics[width=.48\textwidth]{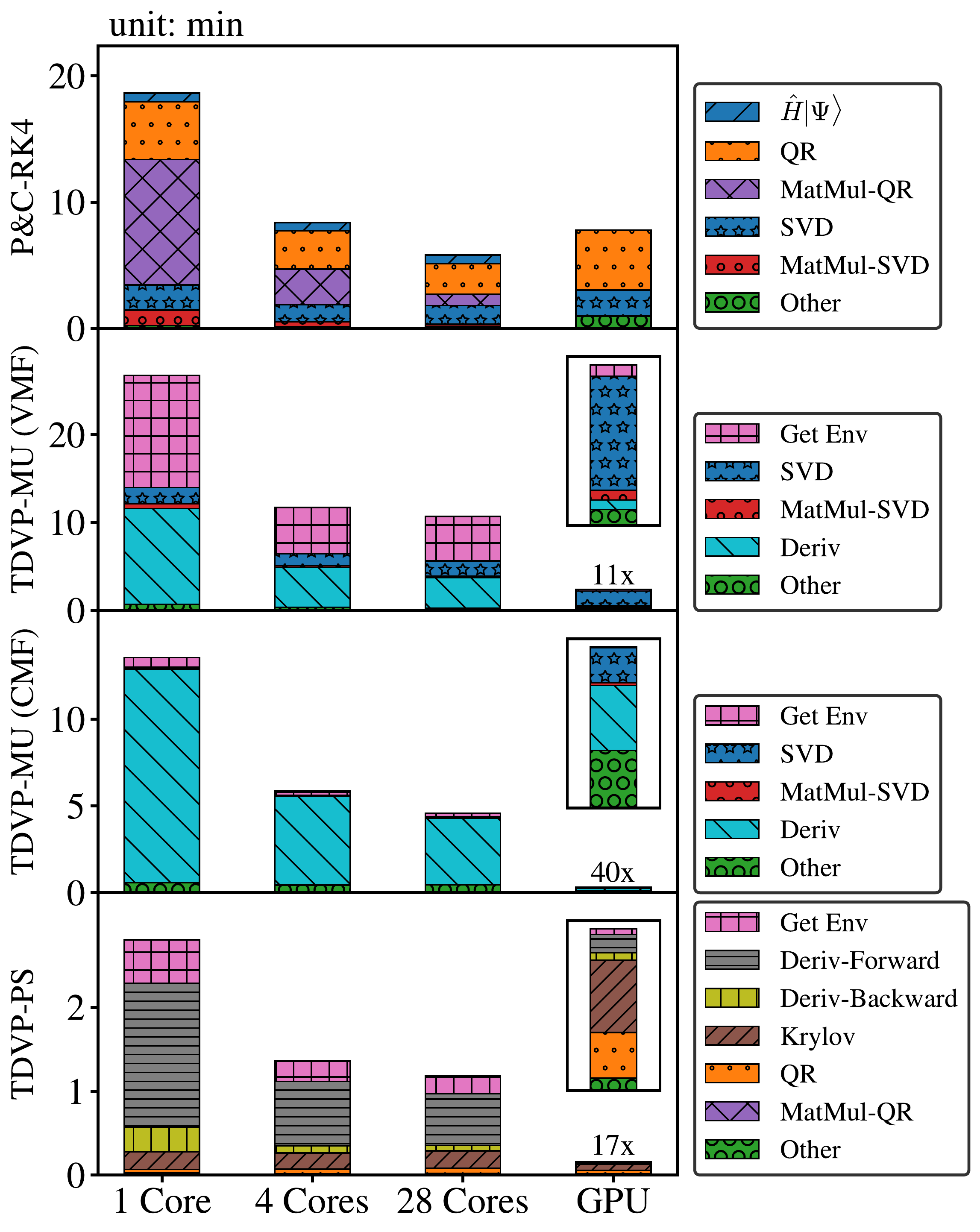}
  \caption{The time cost of a single evolution step and its intensive sub-steps with $M_\textrm{S}=128$ and $\tau=160$.
  The bars for CPU-GPU heterogeneous computation are further shown in the insets for clarity. (See the text in Section~\ref{sec:methods} for the definitions and labels of the sub-steps.)}
  \label{fig:time-cost-128}
\end{figure}

For the three schemes discussed in this paper with $M_\textrm{S}=128$,
the wall time cost of a single evolution step ($\tau=160$) as well as the composed sub-steps 
are presented in Fig.~\ref{fig:time-cost-128}. For TDVP-MU (VMF), there are several steps within $\tau$.
We firstly focus on the total time cost of each scheme in Fig.~\ref{fig:time-cost-128} from which
we can learn that for all schemes the parallelization implemented in the standard linear algebra library on CPU is not optimal for TD-DMRG. 
Using 4 cores can merely double the computational speed, 
and further improvement is not significant up to using all 28 cores of the CPU. 
A more inspiring fact is that the CPU-GPU heterogeneous computing is able to boost the speed of time evolution up to 40 times for TDVP
based algorithms, although for  P\&C-RK4  the effect of the GPU is comparable with 4 cores of CPU.
We note that when doing calculations on GPU, the GPU usage is not always 100\%, so 
more drastic speedup is expected with even larger bond dimension. 
Indeed, when $M_\textrm{S}=256$ the time costs of  TDVP-MU (CMF)  on single core CPU and the heterogeneous CPU-GPU are 4572 s and 63 s respectively,
indicating a 73-fold acceleration.

A closer inspection of the sub-steps shown in Fig.~\ref{fig:time-cost-128} reveals more details about the acceleration. For  P\&C-RK4 , the bottlenecks of the algorithm are ``QR'' and ``MatMul-QR''  as
discussed in Section~\ref{sec:p&C}. Although ``MatMul-QR'' can be efficiently accelerated by multi-core
CPU and CPU-GPU heterogeneous computing, the same is not true for ``QR'' and ``SVD'',
limiting the overall acceleration efficiency.
When the calculation runs on the GPU, the matrix multiplication virtually costs no time while 
``QR''  and ``SVD'' become the bottlenecks.
When the GPU is incoperated in  TDVP-MU (VMF) , the SVD decomposition for regularization becomes prominent even
though it only scales at $\order{NM_\textrm{S}^3d}$ in a single step. 
Thus, it is suggested to return back to the original equation in Eq.~\ref{eq:tdvp1_L} for higher efficiency
once the regularization does not take effect anymore, 
owing to the fact that calculating the inverse of $S$ directly only costs $\order{NM_\textrm{S}^3}$.
Even better acceleration efficiency can be achieved in  TDVP-MU (CMF) 
because the SVD decomposition is carried out less frequently.
The ``Get Env'' part also takes less time in CMF than VMF due to the same reason.
The time costs of bottleneck sub-steps of  TDVP-PS  are similar with that of  TDVP-MU (CMF) 
except that the integrator seems to take a large fraction of time especially when running on the GPU.
This is considered to be a demonstration of the GPU latency
which makes the total time cost of lots of operations on small matrices 
larger than the total time cost of a few operations on large matrices.
When $M_\textrm{S}$ is increased to 256 the relative time cost of the ``Krylov'' part is diminished.
The exact abnormality with GPU acceleration is also manifested to a lesser extent in  TDVP-MU (CMF) 
where the time-consuming ``other'' part is primarily composed of operations in the Runge-Kutta integrator. 
Hence, if a smaller bond dimension is employed, devoting more resource on the computation
will not gain much benefit.
In fact, for TDVP based schemes the absolute wall time with GPU are roughly the same between $M_\textrm{S}=128$ and $M_\textrm{S}=32$.
For example, the time costs for a single step of evolution
in  TDVP-PS  with $M_\textrm{S}=128$ and $M_\textrm{S}=32$ are 10.8 s and 7.9 s respectively.
This phenomenon implies that when $M_\textrm{S} \le 128$ the latency of GPU calculation is of the same order with the actual computational time cost. Therefore, with the CPU-GPU heterogeneous computation, TD-DMRG simulation with $M_\textrm{S} > 100$  for a large-scale system would become a routine task in the future.


  
\section{Conclusion}

To summarize, in this paper, we carry out numerical benchmarks on three different TD-DMRG time evolution schemes,
which are P\&C-RK4, TDVP-MU and TDVP-PS, in terms of accuracy based on the vibronic coupling model represented by the FMO complex.
After defining the tangent space projector for a general non-canonical MPS, the EOMs of TDVP-MU and TDVP-PS are re-derived from the same starting point but with different gauge conditions. 
The numerical results demonstrate that TDVP-MU and TDVP-PS could indeed obtain similar accuracy with a converged time step size $\dt$ , while P\&C-RK4 is not as accurate as them. However, surprisingly, the converged time step size of TDVP-PS is at least 16 times larger than that of TDVP-MU (CMF), although both of them implemented in this work are second order methods.
Regarding computational efficiency, 
we firstly analyze the complexity of each evolution scheme. To further accelerate the intensive tensor computations in TD-DMRG, we adopt a CPU-GPU heterogeneous computing strategy, 
in which CPU and GPU are respectively responsible for the tensor decomposition and contraction. 
We find that for the TDVP-based schemes where the tensor contraction is the main bottleneck, 
this heterogeneous computing approach is able to speed up these two schemes by up to 73 times ($M_\textrm{S}=256$). 
Finally, we conclude that with the accurate and efficient TDVP-based evolution schemes (especially TDVP-PS) and CPU-GPU heterogeneous algorithm/hardware, TD-DMRG has been a promising method for the large-scale quantum dynamics simulation of real chemical and physical problems.

\begin{acknowledgements}
This work is supported by the National Natural Science Foundation of China through the project ``Science CEnter for Luminescence from Molecular Aggregates (SCELMA)'' Grant Number 21788102, as well as by the Ministry of Science and Technology of China through the National Key R\&D Plan Grant Number 2017YFA0204501. J.R. is also supported by the Shuimu Tsinghua Scholar Program.
The authors are indebted to Prof.  Garnet Chan for the stimulating discussions.
The authors also gratefully thank Mr. Hongde Yu for support on crystal structure of the FMO complex. 
\end{acknowledgements}

\appendix
\section{The validity of the reference state}
\label{sec:app-valid}

Here we present the data for the numerical convergence of the reference
obtained by  TDVP-PS  with $M_\textrm{S}=256$ and $\dt=10$.
The mean cumulative deviation with various $M_\textrm{S}$ and $\dt=10$
at $t_1=6560$ and $t_2=41120$
is shown in Table~\ref{tab:convergence-m},
while the same data for $M_\textrm{S}=256$ and various $\dt$
is shown in Table~\ref{tab:convergence-dt}.
\begin{table}
\caption{\label{tab:convergence-m}
The mean cumulative deviation with various $M_\textrm{S}$
and $\dt=10$ by  TDVP-PS 
at $t_1=6560$ and $t_2=41120$. (Reference: $M_\textrm{S}=256, \tau=10$)}
\begin{ruledtabular}
\begin{tabular}{ccc}
      & \multicolumn{2}{c}{error ($t$)} \\
$M_\textrm{S}$   & $t_1=6560$       & $t
_2=41120$ \\
\hline
32  & $6 \times 10^{-7}$    & $1 \times 10^{-3}$     \\
64  & $8 \times 10^{-9}$    & $3 \times 10^{-4}$    \\
128 & $3 \times 10^{-10}$   & $3 \times 10^{-5}$   \\
192 &  $3 \times 10^{-11}$  & $7 \times 10^{-6}$
\end{tabular}
\end{ruledtabular}
\end{table}

\begin{table}
\caption{\label{tab:convergence-dt}
The mean cumulative deviation with $M_\textrm{S}=256$
and various $\dt$ by  TDVP-PS 
at $t_1=6560$ and $t_2=41120$. (Reference: $M_\textrm{S}=256, \tau=10$)}
\begin{ruledtabular}
\begin{tabular}{ccc}
      & \multicolumn{2}{c}{error ($t$)} \\
$\dt$   & $t_1=6560$       & $t_2=41120$ \\
\hline
320  & $4 \times 10^{-10}$    & $8 \times 10^{-7}$     \\
160  & $7 \times 10^{-11}$    & $7 \times 10^{-7}$    \\
80   & $9 \times 10^{-13}$    & $8 \times 10^{-7}$   \\
40   & $7 \times 10^{-13}$    & $6 \times 10^{-7}$
\end{tabular}
\end{ruledtabular}
\end{table}

From Table~\ref{tab:convergence-m} and Table~\ref{tab:convergence-dt}
we can see that the reference is accurate
to a level of approximately $10^{-11}$
for short time scale ($t_1=6560$) and $10^{-6}$ for long time scale ($t_2=41120$),
which is considered to be accurate enough for the error scales discussed
in the main text.


\bibliography{reference1.bib}
\end{document}